\documentclass[twocolumn,prb,showpacs]{revtex4}
\usepackage{graphicx}
\usepackage{rotating}
\begin{document}
\title{ Theory for superconductivity in a magnetic field:  A local approximation approach }
\author{Zhidong Hao}
\affiliation{ Department of Physics, University of Science and Technology of China, Hefei, Anhui 230026, China }
\date{ \today }
\begin{abstract}

We present a microscopic theory for superconductivity in a magnetic field based on a local approximation approach. We
derive an expression for free energy density $F$ as a function of temperature $T$ and vector potential {\bf a}, and
two basic equations of the theory: the first is an implicit solution for energy gap parameter amplitude $|\Delta_{\bf
k}|$ as a function of wave vector {\bf k}, temperature $T$ and vector potential {\bf a}; and the second is a
London-like relation between electrical current density {\bf j} and vector potential {\bf a}, with an ``effective
superconducting electron density'' $n_s$ that is both $T$- and {\bf a}-dependent. The two equations allow
determination of spatial variations of {\bf a} and $|\Delta_{\bf k}|$ in a superconductor for given temperature $T$,
applied magnetic field ${\bf H}_a$ and sample geometry. The theory shows the existence of a ``partly-paired state,''
in which paired electrons (having $|\Delta_{\bf k}|>0$) and de-paired electrons (having $|\Delta_{\bf k}|=0$)
co-exist. Such a ``partly-paired state'' exists even at $T=0$ when $H_a$ is above a threshold for a given sample,
giving rise to a non-vanishing Knight shift at $T=0$ for $H_a$ above the threshold. We expect the theory to be valid
for highly-local superconductors for all temperatures and magnetic fields below the superconducting transition. In
the low-field limit, the theory reduces to the local-limit result of BCS.  As examples, we apply the theory to the
case of a semi-infinite superconductor in an applied magnetic field ${\bf H}_a$ parallel to the surface of the
superconductor and the case of an isolated vortex in an infinite superconductor, and determine, in each case, spatial
variations of quantities such as {\bf a} and $|\Delta_{\bf k}|$.  We also calculate magnetic field penetration depth
$\lambda(T,H_a)$ and lower critical magnetic field $H_{c1}(T)$. The ratio $H_{c1}(T)/H_c(T)$ (where $H_c$ is the
thermodynamic critical magnetic field) is found to be only weakly $T$-dependent for low temperatures and nearly
$T$-independent for intermediate and high temperatures, and quantitatively not very different from that of the
Ginzburg-Landau theory for Ginzburg-Landau parameter $\kappa\gg 1$.

\end{abstract}

\maketitle

\section{ Introduction }

Currently accepted microscopic theories for superconductivity in a magnetic field include the microscopic derivation
of the Meissner effect in the BCS theory\cite{bcs} and the microscopic derivation of the phenomenological
Ginzburg-Landau equations by Gorkov\cite{gorkov} from the BCS theory. However, the BCS derivation of the Meissner
effect is based on a linear-response approach (i.e., an externally applied magnetic field is treated as a weak
perturbation), and thus, is valid only in the low-field limit. The Gorkov derivation of the Ginzburg-Landau equations
is based on the assumption that energy gap function $\Delta({\bf x})$ is a small quantity, and thus, is valid only
for temperatures just below critical temperature $T_c$.  Various extensions of the Gorkov theory also rely heavily on
the assumption of $\Delta({\bf x})$ being small, or the assumption of magnetic field being weak.\cite{gorkovExt}

It is desirable to have a microscopic theory for superconductivity in a magnetic field that is valid under more
general conditions. As an effort along this line, we present in this paper a theory based on a local approximation
approach. The theory allows microscopic description of the suppression of superconductivity by an externally applied
magnetic field. The theory is expected to be valid for highly-local superconductors (the high-temperature
copper-oxide superconductors are examples of highly-local superconductors, for which Ginzburg-Landau parameter
$\kappa\gg 1$) for all temperatures and magnetic fields below the superconducting transition (except, perhaps, in the
high-field limit when the effect of spin paramagnetism, which is neglected in the present work, may become
important). In the low-field limit, the theory reduces to the local-limit result of BCS.\cite{bcs}

In Sec. \ref{secTheory}, we present the details of the theory, and derive an expression for free energy density $F$
as a function of temperature $T$ and vector potential {\bf a}, and two basic equations of the theory: the first is an
implicit solution for energy gap parameter amplitude $|\Delta_{\bf k}|$ as a function of wave vector {\bf k},
temperature $T$ and vector potential {\bf a}; and the second is a relation between electrical current density {\bf j}
and vector potential {\bf a}. We also analyze solutions for $|\Delta_{\bf k}(T,{\bf a})|$. In Sec. \ref{secAppl}, as
examples, we apply the theory to the case of a semi-infinite superconductor in an applied magnetic field ${\bf H}_a$
parallel to the surface of the superconductor and the case of an isolated vortex in an infinite superconductor, and
determine, in each case, spatial variations of quantities such as {\bf a} and $|\Delta_{\bf k}|$. We also calculate
magnetic field penetration depth $\lambda(T,H_a)$ and lower critical magnetic field $H_{c1}(T)$. A brief summary is
given in Sec. \ref{secSummary}.

\section{Theory }
\label{secTheory}

We consider a superconductor in an applied magnetic field. Our starting point is the same BCS pairing
Hamiltonian\cite{bcs}
\begin{equation}
\hat{H}  = \! \sum_{\bf k\sigma} \left(\epsilon_{\bf k} \!-\! \epsilon_{_F}\right) c^{\dagger}_{\bf k\sigma}c_{\bf
k\sigma} \mbox{} + \!\sum_{\bf kk'} V_{\bf kk'} c^{\dagger}_{\bf k\uparrow}c^{\dagger}_{\bf -k\downarrow} c_{- \bf
k'\downarrow}c_{\bf k'\uparrow} , \label{H0}
\end{equation}
where $\epsilon_{\bf k}$ is the normal state single-electron energy, $\epsilon_{_F}$ the Fermi energy, $V_{\bf kk'}$
the pairing interaction matrix element, and $c^\dagger_{{\bf k}\sigma}$ and $c_{{\bf k}\sigma}$ are the Fermi
operators of an electronic state of wave vector {\bf k} and spin $\sigma$ in the normal state. Single-electron energy
$\epsilon_{\bf k}$ and its corresponding single-electron wave function $\phi_{\bf k}({\bf x})$ satisfy the
Schr\"{o}dinger equation
\begin{equation}
\label{SchEq} H({\bf x}) \phi_{\bf k}({\bf x}) = \epsilon_{\bf k}
\phi_{\bf k}({\bf x})
\end{equation}
with single-electron Hamiltonian
\begin{equation}
\label{Hx} H({\bf x}) = \frac{1}{2m}\left[-i\hbar\nabla + \frac{e}{c}{\bf a}({\bf x})\right]^2 + U({\bf x}),
\end{equation}
where ${\bf a}({\bf x})$ is the vector potential, and $U({\bf x})$ a periodic scalar potential.

In writing down Hamiltonian $\hat{H}$, we have neglected, for simplicity, the effect of spin paramagnetism (which may
be important for high fields).

We note that, if $\epsilon_{\bf k}$ and $\phi_{\bf k}({\bf x})$ for ${\bf a}\neq 0$ are known, $\hat{H}$ can be
diagonalized in essentially the same way as for ${\bf a}=0$ (i.e, by making the BCS pairing approximation, and then
applying the Bogoliubov transformation\cite{bogo}).  However, since ${\bf a}({\bf x})$ in the superconducting state
is itself an unknown function, a simultaneous determination of $\epsilon_{\bf k}$, $\phi_{\bf k}({\bf x})$ and ${\bf
a}({\bf x})$ does not seem possible. We therefore adopt a local approximation approach, which we describe next.

\subsection{ Local Approximation Approach }

We note that, in the superconducting state, vector potential ${\bf a}({\bf x})$ varies spatially with the length
$\lambda$, the magnetic field penetration depth, which is $\sim\!10^3\text{\AA}$.  In contrast, single-electron wave
function $\phi_{\bf k}({\bf x})$ oscillates spatially with a much shorter length $k_F^{-1}\!\!\sim\!1\text{\AA}$
(here ${\bf k}_F$ is a Fermi wave vector),\cite{ashcroft} so that ${\bf a}({\bf x})$ can be considered locally
constant over many wavelengths of $\phi_{\bf k}({\bf x})$. Our approach is based on this observation, and can be
outlined as consisting the following three steps.

{\it Step 1:} We focus on a small region around a local point, say ${\bf x'}$. Dimension $D$ of this small region
satisfies $d\ll D\ll \lambda$, where $d\sim 1\text{\AA}$ is a crystal lattice constant of the superconductor.  In
this small region, we assume ${\bf a}({\bf x})={\bf a}({\bf x'})$ is a constant, and solve the Schr\"{o}dinger
equation [Eq. (\ref{SchEq})] to obtain $\epsilon_{\bf k}$ and $\phi_{\bf k}({\bf x})$.

{\it Step 2:} Based on the obtained $\epsilon_{\bf k}$ and $\phi_{\bf k}({\bf x})$ in the vicinity of ${\bf x'}$, we
diagonalize Hamiltonian $\hat{H}$, which now becomes a local quantity, because of its dependence upon ${\bf a}({\bf
x'})$ through $\epsilon_{\bf k}$ and $\phi_{\bf k}({\bf x})$.

{\it Step 3:} Once Hamiltonian $\hat{H}$ is diagonalized, we move on to derive an expression for local free energy
density $F$ and two basic equations of the theory. The equations allow determination of the spatial variations of
vector potential ${\bf a}$ and other quantities for given temperature, applied magnetic field and sample geometry.

Since the non-local effect (or coherence effect)\cite{bcs,pippard} in the superconducting state is not accounted for,
we expect this approach to be valid only for highly-local superconductors for which magnetic field penetration depth
$\lambda$ is much larger than coherence length $\xi$ (i.e., $\lambda \gg \xi$).

We explain the details of this approach in the following subsections.

\subsection{ Determination of $\epsilon_{\bf k}$ and $\phi_{\bf k}({\bf x})$ }

As outlined above, we first focus on a small region around a local point, say ${\bf x'}$.  Dimension $D$ of this
small region satisfies $d \ll D \ll \lambda$, where $d$ is a crystal lattice constant of the superconductor, and
$\lambda$ is the magnetic field penetration depth. Since vector potential ${\bf a}({\bf x})$ varies spatially with
the length $\lambda$, which is $\sim\!10^3\text{\AA}$, whereas $\phi_{\bf k}({\bf x})$ oscillates spatially with the
length $d$, which is $\sim\!1\text{\AA} \ll \lambda$, we can assume ${\bf a}({\bf x})={\bf a}({\bf
x'})=\text{constant}$ in this small region, and solve Eq. (\ref{SchEq}).

For a constant ${\bf a}({\bf x})={\bf a}({\bf x'})$, it is not difficult to solve Eq. (\ref{SchEq}).  For free
electrons, for which scalar potential $U({\bf x})=\text{constant}$, solutions of Eq. (\ref{SchEq}) are easily
obtained.  Namely, for ${\bf a}=0$, we have
\begin{equation}
\epsilon^{(0)}_{\bf k}=\hbar^2k^2/2m
\end{equation}
and
\begin{equation}
\phi^{(0)}_{\bf k}({\bf x})=e^{i{\bf k}\cdot{\bf x}};
\end{equation}
and for ${\bf a}\neq 0$, we have
\begin{equation}
\epsilon_{\bf k} = \left(\hbar{\bf k}+\frac{e}{c}{\bf a}\right)^2/2m  \label{ek}
\end{equation}
and
\begin{equation}
\phi_{\bf k}({\bf x}) = e^{i{\bf k}\cdot{\bf x}}.
\end{equation}
Here we have used $\epsilon^{(0)}_{\bf k}$ and $\phi^{(0)}_{\bf k}$ to denote solutions of Eq. (\ref{SchEq}) for
${\bf a}=0$.

In this paper, we will not consider the case of a general periodic scalar potential $U({\bf x})$ (i.e., we will not
consider in this paper how the details of an electronic energy band structure may affect properties of the
superconducting state). Instead, for simplicity in presenting the theory, we will use the solutions for free
electrons in the following.

\subsection{ Diagonalization of $\hat{H}$ }

Having obtained $\epsilon_{\bf k}$ and $\phi_{\bf k}({\bf x})$, we can move on to diagonalize Hamiltonian $\hat{H}$
of Eq. (\ref{H0}), which now becomes a local quantity, because $\epsilon_{\bf k}$ and $\phi_{\bf k}({\bf x})$ are
obtained locally at ${\bf x'}$ for ${\bf a} = {\bf a}({\bf x'})$.

An important step in the diagonalization of Hamiltonian $\hat{H}$ is assuming that the pairing\cite{bcs} of electrons
of opposite momenta and spins holds even for ${\bf a}\neq 0$. I.e., for a pair of $({\bf k}\!\uparrow)$ and $({\bf
-k}\!\downarrow)$ electrons, we assume
\begin{equation}
\langle c_{-\bf k\downarrow}c_{\bf k\uparrow} \rangle \neq 0  \label{pairing}
\end{equation}
when the electrons are superconducting (here the angle brackets $\langle\cdots\rangle$ denote a thermal average).

The diagonalization of Hamiltonian $\hat{H}$ is the same as in the case of ${\bf a}=0$,\cite{bogo} except that we
have $\epsilon_{-\bf k} \neq \epsilon_{\bf k}$ for ${\bf a} \neq 0$. The results of the diagonalization are as
follows.

The energy gap parameter is defined as\cite{bcs}
\begin{equation}
\Delta_{\bf k} = - \sum_{\bf k'} V_{\bf k,k'} \langle c_{-\bf k'\downarrow}c_{\bf k'\uparrow} \rangle .
\label{gap-def}
\end{equation}

The diagonalized Hamiltonian is
\begin{equation}
{\hat H} = \sum_{\bf k}\left( U_{\bf k} + E_{\bf k} \gamma^{\dagger}_{\bf k\uparrow} \gamma_{\bf k\uparrow}  +
E_{-\bf k} \gamma^{\dagger}_{-\bf k\downarrow} \gamma_{-\bf k\downarrow}\right), \label{H}
\end{equation}
where
\begin{equation}
U_{\bf k} = \frac{ \xi_{\bf k} + \xi_{-\bf k} } { 2 } - E^{(s)}_{\bf k} + \frac{|\Delta_{\bf k}|^2}{2E^{(s)}_{\bf k}}
\left(1-f_{\bf k}-f_{-\bf k}\right); \label{Uk}
\end{equation}
\begin{equation}
\xi_{\bf k} = \epsilon_{\bf k} - \epsilon_{_F}
\label{xik}
\end{equation}
is the single-electron energy in the normal state, measured relative to the Fermi energy $\epsilon_{_F}$;
\begin{equation}
E_{\bf k} = E^{(s)}_{\bf k} + \frac{\xi_{\bf k}-\xi_{-\bf k}}{ 2 } \label{Ek}
\end{equation}
the quasi-particle excitation energy in the superconducting state;
\begin{equation}
E^{(s)}_{\bf k}=\sqrt{\left(\frac{\xi_{\bf k}+\xi_{-\bf k}}{2} \right)^2 + |\Delta_{\bf k}|^2 } \label{Eks}
\end{equation}
the symmetric part of $E_{\bf k}$;
\begin{equation}
f_{\bf k} =  \left(e^{E_{\bf k}/k_BT}+1\right)^{-1}
\end{equation}
the Fermi function; and $\gamma^\dagger_{{\bf k}\sigma}$ and $\gamma_{{\bf k}\sigma}$ are the Fermi operators for
quasi-particles in the superconducting state.

The operators $\gamma^\dagger_{{\bf k}\sigma}$ and $\gamma_{{\bf k}\sigma}$ are related to $c^\dagger_{{\bf
k}\sigma}$ and $c_{{\bf k}\sigma}$ via the Bogoliubov transformation\cite{bogo}
\begin{equation}
\left(
\begin{array}{c}
c_{\bf k\uparrow} \\ c^{\dagger}_{-\bf k\downarrow}
\end{array}
\right) = \left(
\begin{array}{cc}
u^{\star}_{\bf k}   &  v_{\bf k} \\
- v^{\star}_{\bf k}  &  u_{\bf k}
\end{array}
\right) \left(
\begin{array}{c}
\gamma_{\bf k\uparrow} \\  \gamma^{\dagger}_{-\bf k\downarrow}
\end{array}
\right),
\end{equation}
where the coefficients $u_{\bf k}$ and $v_{\bf k}$ satisfy the following relations:
\begin{equation}
|u_{\bf k}|^2 = \frac{1}{2} \left( 1 + \frac{ \xi_{\bf k} + \xi_{-\bf k} } { 2 E^{(s)}_{\bf k} } \right),
\end{equation}
\begin{equation}
|v_{\bf k}|^2 = \frac{1}{2} \left( 1 - \frac{ \xi_{\bf k} + \xi_{-\bf k} } { 2 E^{(s)}_{\bf k} } \right),
\end{equation}
and
\begin{equation}
\Delta_{\bf k}u_{\bf k}v^{\star}_{\bf k} = \frac{ |\Delta_{\bf k}|^2 } { 2 E^{(s)}_{\bf k} }.
\end{equation}

After the diagonalization of Hamiltonian $\hat{H}$, Eq. (\ref{gap-def}) can be expressed as
\begin{equation}
\Delta_{\bf k} = \mbox{} - \sum_{\bf k'} V_{\bf k,k'} \frac{1-f_{\bf k'}-f_{-\bf k'} } { 2 E^{(s)}_{\bf k'} }
\Delta_{\bf k'}. \label{gap-eq}
\end{equation}

With regard to the above-described diagonalization of Hamiltonian $\hat{H}$, the following are worth emphasizing.

(i) Interaction matrix element $V_{\bf k,k'}$ is ${\bf a}$-independent, because $\phi_{\bf k}({\bf x}) =
\phi^{(0)}_{\bf k}({\bf x})$.

(ii) For ${\bf a}\neq 0$, since $\epsilon_{-\bf k} \neq \epsilon_{\bf k}$, we have $\xi_{-\bf k}\neq\xi_{\bf k}$,
$E_{-\bf k} \neq E_{\bf k}$ and $f_{-\bf k} \neq f_{\bf k}$. However, we still have $U_{-\bf k}=U_{\bf k}$,
$E^{(s)}_{-\bf k}=E^{(s)}_{\bf k}$ and $\Delta_{-\bf k} = \Delta_{\bf k}$, as one can see from Eqs. (\ref{Uk}),
(\ref{Eks}) and (\ref{gap-eq}).

(iii) Fermi energy $\epsilon_{_F}$, relative to which energies such as $\xi_{\bf k}$ and $E_{\bf k}$ are measured, is
a local quantity, i.e., $\epsilon_{_F} = \epsilon_{_F}({\bf a}({\bf x'}))$.  This can be understood as follows.  In
the superconducting ground state, all electrons are paired.  For each pair of electrons, we have $(\epsilon_{\bf
k}+\epsilon_{-\bf k})\neq 2\epsilon^{(0)}_{\bf k}$ for ${\bf a}\neq 0$.  Thus, we expect $\epsilon_{_F} \neq
\epsilon^{(0)}_{_F}$ for ${\bf a}\neq 0$.  We further expect that $\epsilon_{_F} > \epsilon^{(0)}_{_F}$ for ${\bf
a}\neq 0$ in the superconducting state, where the increase in the Fermi energy is due to a flow of the paired
electrons.

For an electronic energy spectrum of the form of Eq. (\ref{ek}), Fermi energy $\epsilon_{_F}$ is determined by
$\epsilon_{_F}\!=\!(\epsilon_{{\bf k}_{_F}}+\epsilon_{{-\bf k}_{_F}})/2$, which gives
\begin{equation}
\epsilon_{_F}=\epsilon^{(0)}_{_F}+\frac{e^2}{2mc^2}a^2. \label{eF}
\end{equation}

With the help of Eqs. (\ref{ek}) and (\ref{eF}), we also have
\begin{equation}
\frac{\xi_{\bf k}+\xi_{-\bf k}}{2}=\xi^{(0)}_{\bf k} \label{xi+xi}
\end{equation}
and
\begin{equation}
\frac{\xi_{\bf k}-\xi_{-\bf k}}{2}=\frac{\hbar e}{mc} {\bf k}\cdot{\bf a}\,, \label{xi-xi}
\end{equation}
where
\begin{equation}
\xi^{(0)}_{\bf k}=\epsilon^{(0)}_{\bf k}-\,\,\epsilon^{(0)}_{_F} .
\end{equation}

From the above-described diagonalization of Hamiltonian $\hat{H}$, we see the following: Electronic states (or
quasi-particles) in the superconducting state are each characterized by a wave-vector {\bf k} and a spin $\sigma$,
and are in one-to-one correspondence with those in the normal state (this point is the same as in the case of ${\bf
a}=0$).\cite{bcs,bogo} The superconducting state is different from the normal state because (i) there exists an
energy gap parameter $\Delta_{\bf k}$ for an electronic excitation in the superconducting state; and (ii) Fermi
energy $\epsilon_{_F}$ is vector potential {\bf a} dependent in the superconducting state. Both the existence of
$\Delta_{\bf k}$ and the {\bf a}-dependence of $\epsilon_{_F}$ originate from the pairing of $({\bf k}\!\uparrow)$
and $({\bf -k}\!\downarrow)$ electrons.

\subsection{ Free energy density $F$ }


From diagonalized Hamiltonian $\hat{H}$ [Eq. (\ref{H})], the following expression for free energy density $F$ at
location {\bf x} (we can drop the prime in ${\bf x'}$ hereafter) in the superconductor can be derived:
\begin{eqnarray}
F &\!\!\! = &\!\!\!\sum_{\bf k}\!\left[ U_{\bf k} \!-\! k_BT \ln \!\left( 1 \!+\! e^{-E_{\bf k}/k_BT} \!\right)\!\!
\left( 1
\!+\! e^{-E_{-\bf k}/k_BT}\!\right) \!\right] \nonumber \\
& & \mbox{} + \,\,\, \frac{ne^2}{2mc^2}a^2 \,\,\, + \,\,\, \frac{ (\nabla\!\times\!{\bf a})^2 } { 8 \pi} \, .
\label{F}
\end{eqnarray}

The first of the three terms in the above expression comes from $\,\,-k_BT\ln \left[ \text{Tr} \left(
e^{-\hat{H}/k_BT} \right) \right]$, which is the usual statistical electronic free energy density.\cite{feynman}  The
second term comes from $\,n(\epsilon_{_F} \!-\! \epsilon^{(0)}_{_F})$, where $n$ is the density of electrons, and we
have used Eq. (\ref{eF}).  This term is added to $F$ because electronic energies in the expression for Hamiltonian
$\hat{H}$ are measured relative to $\epsilon_{_F}$. The third term is the magnetic field energy density.

Note that the expression for $F$ involves $|\Delta_{\bf k}|$ (through $U_{\bf k}$, $E_{\bf k}$ and $E_{-\bf k}$). As
we will see in the next subsection, $|\Delta_{\bf k}|$ is a function of {\bf k}, $T$ and {\bf a}, i.e., $|\Delta_{\bf
k}|=|\Delta_{\bf k}(T,{\bf a})|$, and the function $|\Delta_{\bf k}(T,{\bf a})|$ is determined by the
self-consistency of the diagonalization of $\hat{H}$.  Thus, we see that $F$ is a function of  $T$ and {\bf a}, i.e.,
$F=F(T,{\bf a})$.  [As can be shown, free energy density $F$, excluding the magnetic field energy density, becomes
{\bf a}-independent when $|\Delta_{\bf k}|=0$ for all electronic states.]

\subsection{ First equation: Implicit solution for $|\Delta_{\bf k}(T,{\bf a})|$ }

The following equation is derived as a condition for the self-consistency of the diagonalization of Hamiltonian
$\hat{H}$:
\begin{equation}
\frac{ 1 - f_{\bf k} - f_{-\bf k} } { E^{(s)}_{\bf k} } = \text{ independent of $T$ and {\bf a}\,. } \label{haoEq}
\end{equation}

For ${\bf a}=0$, this equation was previously derived by the author in Ref. \onlinecite{hao93}. We present the
details of the derivation of this equation for ${\bf a}\neq 0$, which is similar to that for ${\bf a}= 0$, in
Appendix \ref{deriveHaoEq}.

By using $1-2f_{\bf k} = \tanh(E_{\bf k}/2k_BT)$ and the condition that $|\Delta_{\bf k}| = 0$ at $(T,\,{\bf a}) =
(T_c,0)$, we can express the above equation as
\begin{widetext}
\[
\left[2 \sqrt{ \left( \xi^{(0)}_{\bf k} \right)^2 + |\Delta_{\bf k}|^2 } \,\, \right]^{-1} \left\{ \tanh \left[
\left( \sqrt{ \left( \xi^{(0)}_{\bf k} \right)^2 + |\Delta_{\bf k}|^2 } + \frac{\hbar e}{mc}{\bf k}\!\cdot{\bf a}
\,\, \right) / 2k_BT \right] \right.
\]
\begin{equation}
\mbox{} + \left. \tanh \left[ \left( \sqrt{ \left( \xi^{(0)}_{\bf k} \right)^2 + |\Delta_{\bf k}|^2 } - \frac{\hbar
e}{mc}{\bf k}\!\cdot{\bf a} \,\, \right) / 2k_BT \right] \right\} = \frac{ \tanh \left( |\xi^{(0)}_{\bf k}| / 2k_BT_c
\right) } { |\xi^{(0)}_{\bf k}| } \,. \label{HaoEq}
\end{equation}
\end{widetext}

This equation is an implicit solution for $|\Delta_{\bf k}|$ as a function of {\bf k}, $T$ and {\bf a} for given
$T_c$.

Note that interaction $V_{\bf k,k'}$ does not appear in Eq. (\ref{HaoEq}).  Instead, critical temperature $T_c$ is
involved through the condition that $|\Delta_{\bf k}|=0$ at $(T,\,{\bf a})=(T_c,0)$.  Namely, $|\Delta_{\bf k}|$
depends on $V_{\bf k,k'}$ only through $T_c$.

Critical temperature $T_c$ and phase $\theta_{\bf k}$ of $\Delta_{\bf k}$ (i.e., $\Delta_{\bf k}=|\Delta_{\bf
k}|e^{i\theta_{\bf k}}$) are determined by solving the eigenvalue problem
\begin{equation}
\Delta_{\bf k} = - \sum_{{\bf k}'} V_{{\bf k},{\bf k}'} \frac{\tanh(|\xi^{(0)}_{\bf k'}|/2k_BT_c)}{2|\xi^{(0)}_{\bf
k'}|} \Delta_{{\bf k}'} \label{tc-eq}
\end{equation}
for given interaction $V_{\bf k,k'}$ and electronic energy spectrum $\xi^{(0)}_{\bf k}$.  Equation (\ref{tc-eq}) is
the linearized form of Eq. (\ref{gap-eq}) in the limit of $(T,\,{\bf a}) \rightarrow (T_c,0)$.  A complete solution
for $\Delta_{\bf k}$ is therefore a combination of the solution for $|\Delta_{\bf k}|$ of Eq. (\ref{HaoEq}) and the
solutions for $T_c$ and $\theta_{\bf k}$ of Eq. (\ref{tc-eq}). (Related discussions are also given in Refs.
\onlinecite{hao96} and \onlinecite{hao}.)

\subsection{ Second equation: Relation between ${\bf j}$ and ${\bf a}$ }

Having obtained an expression for $F(T,{\bf a})$ [i.e., the expression given by Eq. (\ref{F}), with $|\Delta_{\bf
k}(T,{\bf a})|$ being implicitly given by Eq. (\ref{HaoEq})], we are now ready to consider determination of ${\bf
a}({\bf x})$. In thermodynamic equilibrium, the overall free energy, given by the volume integral of $F\left(T,{\bf
a}({\bf x})\right)$, must be stationary with respect to arbitrary variation of ${\bf a}({\bf x})$. This variational
problem leads to
\begin{equation}
{\bf j} = -\frac{ne^2}{mc}{\bf a} + \frac{\hbar e}{m}\sum_{\bf k} \left( f_{-\bf k} - f_{\bf k} \right) {\bf k},
\label{HaoEq2}
\end{equation}
where {\bf j} is the electrical current density, and we have used the relations $\frac{4\pi}{c}{\bf j} =
\nabla\!\times\!{\bf b} = \nabla\!\times\!\nabla\!\times\!{\bf a}$\,, with ${\bf b} = \nabla\!\times\!{\bf a}$ being
the magnetic flux density.

The first term on the right-hand side of Eq. (\ref{HaoEq2}) represents a uniform flow of all the electrons, which
originates from the pairing of $({\bf k}\!\uparrow)$ and $({\bf -k}\!\downarrow)$ electrons. The second term is due
to quasi-particle excitations and de-paired electrons, and tends to cancel the first term. When all the electrons are
in the superconducting ground state, the second term is zero.  On the other hand, when $|\Delta_{\bf k}|=0$ for all
the electrons (i.e., when the superconductor is in the normal state), the two terms cancel each other, and we have
${\bf j}=0$.

Equation (\ref{HaoEq2}) was also derived by BCS.\cite{bcs} However, the BCS derivation of Eq. (\ref{HaoEq2}) is based
on a linear-response approach (i.e., vector potential {\bf a} is treated as a small perturbation), and thus, is valid
only in the low-field limit.

Note that Eq. (\ref{HaoEq2}) is not gauge invariant (this feature is the same for the London equation\cite{london}).
This is because the pairing of $({\bf k}\!\uparrow)$ and $({\bf -k}\!\downarrow)$ electrons fixes the total
(canonical) momentum of the electrons, and thereby also the gauge of the vector potential. Namely, the theory became
not-gauge-invariant at the point when the pairing assumption [Eq. (\ref{pairing})] was made. For the same reason, the
expression for $F(T,{\bf a})$ [Eq. (\ref{F})] and the implicit solution for $|\Delta_{\bf k}(T,{\bf a})|$ [Eq.
(\ref{HaoEq})] also are not gauge invariant. Since, as can be shown, $\partial F /\partial a > 0$ (which means that a
larger value of $a$ is energetically less favorable), we see that the gauge of vector potential {\bf a} in the
present theory is such that ${\bf a} \rightarrow  0$ deep inside a bulk superconductor, i.e., the London
gauge.\cite{london}

\begin{table}
\caption{ \label{table1} Units for forming dimensionless quantities  }
\begin{ruledtabular}
\begin{tabular}{ll}
Quantity & Unit \\
\hline
Length & $\lambda_0$ \\
Temperature & $T_c$ \\
Energy & $k_BT_c$ \\
Magnetic field & $H_{c0}$ \\
Vector potential & $\lambda_0H_{c0}$ \\
Electrical current density & $(c/4\pi)H_{c0}/\lambda_0$ \\
Energy density & $H_{c0}^2/8\pi$ \\
Effective superconducting electron density & $n$ \\
\end{tabular}
\end{ruledtabular}
\end{table}

\subsection{ Dimensionless units }

It is convenient to introduce a set of units so that physical quantities involved in the theory become dimensionless.
The units that we choose to use are listed in Table 1, where $\lambda_0$ is the magnetic field penetration depth at
$T=0$ in the limit of zero magnetic field, and satisfies $\lambda_0^2=mc^2/4\pi ne^2$; and $H_{c0}$ is the
thermodynamic critical magnetic field at $T=0$, and satisfies $H_{c0}^2/8\pi=(\pi^2/6)(k_BT_c)^2N_0$, where $N_0$ is
the density of states at the Fermi level ($N_0=mk_{_F}/2\pi^2\hbar^2$ for free electrons\cite{ashcroft}).  The result
for $H_{c0}$ was previously obtained by the author in Ref. \onlinecite{hao96}.

We also make the substitution $\sum_{\bf k} \rightarrow (1/8\pi^3)\!\int \!d^3{\bf k}$, and, as usual, assume that
$\epsilon^{(0)}_{_F}\gg k_BT_c$ so that the substitution $\int_{0}^{\infty} d\epsilon^{(0)} \rightarrow
\int_{-\infty}^{\infty} d\xi^{(0)}$ and the approximation ${\bf k}\cdot {\bf a} = kaz_{\bf k}\simeq k_{_F}az_{\bf k}$
hold, where $z_{\bf k}=\cos\alpha_{\bf k}$, with $\alpha_{\bf k}$ being the angle between {\bf k} and {\bf a}.

By using the units shown in Table 1 and the above-mentioned assumption and approximation, Eq. (\ref{F}) becomes
\begin{widetext}
\begin{equation}
F = \frac{6}{\pi^2} \int_{-\infty}^{\infty} d\xi^{(0)}_{\bf k}\int_0^1 dz_{\bf k} \left[ U_{\bf k} - T \ln \left( 1
\!+\! e^{-E_{\bf k}/T} \right) \left( 1 + e^{-E_{-\bf k}/T}\right) \right] + a^2 + (\nabla\!\times\!{\bf a})^2
\label{f}
\end{equation}
with
\[
U_{\bf k} = \xi_{\bf k}^{(0)} - \sqrt{ \left( \xi^{(0)}_{\bf k} \right)^2 + |\Delta_{\bf k}|^2 } + |\Delta_{\bf
k}|^2\frac{\tanh\left(|\xi_{\bf k}^{(0)}|/2\right)}{2|\xi_{\bf k}^{(0)}|}
\]
and
\[
E_{\pm\bf k} = \sqrt{ \left( \xi^{(0)}_{\bf k} \right)^2 + |\Delta_{\bf k}|^2 } \pm \frac{\pi}{\sqrt{2}} az_{\bf k}
\,;
\]
and Eqs. (\ref{HaoEq}) and (\ref{HaoEq2}) become
\[
\left[2 \sqrt{ \left( \xi^{(0)}_{\bf k} \right)^2 + |\Delta_{\bf k}|^2 } \,\, \right]^{-1} \left\{ \tanh \left[
\left( \sqrt{ \left( \xi^{(0)}_{\bf k} \right)^2 + |\Delta_{\bf k}|^2 } + \frac{\pi}{\sqrt{2}} az_{\bf k} \,\,
\right) / 2T \right] \right.
\]
\begin{equation}
\mbox{} + \left. \tanh \left[ \left( \sqrt{ \left( \xi^{(0)}_{\bf k} \right)^2 + |\Delta_{\bf k}|^2 } -
\frac{\pi}{\sqrt{2}} az_{\bf k} \,\, \right) / 2T \right] \right\} = \frac{ \tanh \left( |\xi^{(0)}_{\bf k}| / 2
\right) } { |\xi^{(0)}_{\bf k}| } \label{haoeq1}
\end{equation}
\end{widetext}
and
\begin{equation}
\label{haoeq2} {\bf j} = -n_s{\bf a} \, ,
\end{equation}
respectively, where the ``effective superconducting electron density''
\begin{equation}
\label{ns} n_s = 1 - \frac{3\sqrt{2}}{\pi a} \int_{0}^{\infty} \!d\xi^{(0)}_{\bf k} \int_{0}^{1} \!dz_{\bf k} \,
z_{\bf k} \left( f_{-\bf k} - f_{\bf k} \right) \label{ns-eq}
\end{equation}
with
\begin{equation}
f_{\pm\bf k} = \left[ e^{ \left( \sqrt{ \left( \xi^{(0)}_{\bf k} \right)^2 + |\Delta_{\bf k}|^2 } \pm
\frac{\pi}{\sqrt{2}} az_{\bf k} \right)/T }  + 1  \right]^{-1} \!\!\!.
\end{equation}

Note that Eq. (\ref{haoeq2}) is the same as the London equation,\cite{london} except that our $n_s$, given by Eq.
(\ref{ns-eq}), is both $T$- and {\bf a}-dependent, whereas $n_s$ is only $T$-dependent in the London theory.

The condensation energy density is defined as
\begin{equation}
F_c = F'_n - F'_s\,, \label{Fc}
\end{equation}
where $F'_s = F - (\nabla\times{\bf a})^2$ is the superconducting state free energy density excluding the magnetic
field energy density; and $F'_n$ is the normal state counterpart of $F'_s$. The quantity $F_c$ can be used as a
measure of the difference between the normal and superconducting states.

As can be shown, the following relation exists:
\begin{eqnarray}
n_s & = & \,\,\, \frac{1}{2a}\frac{\partial F'_s}{\partial a} \\
    & = & - \frac{1}{2a}\frac{\partial F_c}{\partial a },  \label{nsFc}
\end{eqnarray}
where the second expression holds because $F'_n$ is {\bf a}-independent.

For an isotropic superconductor (as in the present case), $n_s$ and $F_c$ are functions of temperature $T$ and the
magnitude of vector potential {\bf a}, i.e., $n_s=n_s(T,a)$ and $F_c=F_c(T,a)$.

We analyze the functions $|\Delta_{\bf k}(T,{\bf a})|$, $n_s(T,a)$ and $F_c(T,a)$ in the next subsection.

\subsection{ $|\Delta_{\bf k}(T,{\bf a})|$, $n_s(T,a)$, and $F_c(T,a)$}
\label{secDelta}

We solve Eq. (\ref{haoeq1}) to obtain $|\Delta_{\bf k}(T,{\bf a})|$ by using an iterative method.\cite{conte80} [Note
that the variables $({\bf k},\,\, T,\,\, {\bf a})$ for the function $|\Delta_{\bf k}(T,{\bf a})|$ appear in Eq.
(\ref{haoeq1}) in the forms of $(|\xi_{\bf k}^{(0)}|,\,\, T,\,\, az_{\bf k})$; and remember that $az_{\bf
k}=a\cos\alpha_{\bf k}$ is the component of {\bf a} along {\bf k}\,.]  When solving Eq. (\ref{haoeq1}), it is
important to note the following:

(i) We define temperature $T^{\star}_{c{\bf k}}$ as such that Eq. (\ref{haoeq1}) has no $|\Delta_{\bf k}| > 0$
solution for given $|\xi_{\bf k}^{(0)}|$, $T$ and $az_{\bf k}$ if $T>T^{\star}_{c{\bf k}}$.  Note that
$T^{\star}_{c{\bf k}}$ is a function of $|\xi_{\bf k}^{(0)}|$ and $az_{\bf k}$.  Only for ${\bf a}=0$ is
$T^{\star}_{c{\bf k}} = T_c$ the same for all the electronic states.

(ii) Similarly, we define vector potential magnitude $a^{\star}_{c{\bf k}}$ as such that Eq.(\ref{haoeq1}) has no
$|\Delta_{\bf k}| > 0$ solution for given $|\xi_{\bf k}^{(0)}|$, $T$ and $az_{\bf k}$ if $a > a^{\star}_{c{\bf k}}$.
Note that $a^{\star}_{c{\bf k}}$ is a function of $|\xi_{\bf k}^{(0)}|$, $T$ and $z_{\bf k}$.

(iii) Depending on $|\xi_{\bf k}^{(0)}|$, $T$ and $az_{\bf k}$, a $|\Delta_{\bf k}|>0$ solution of Eq. (\ref{haoeq1})
may be an unstable solution, because the normal state solution $|\Delta_{\bf k}| = 0$ may be energetically more
favorable.

Point (iii) can be understood as follows. Note that for given $|\xi_{\bf k}^{(0)}|$, $T$ and $az_{\bf k}$, there are
always two possible solutions for $|\Delta_{\bf k}|$ if $T<T^{\star}_{c{\bf k}}$ or $a<a^{\star}_{c{\bf k}}$: the
$|\Delta_{\bf k}|>0$ solution of Eq. (\ref{haoeq1}) and the $|\Delta_{\bf k}|=0$ solution.  The free energy
associated with a pair of $({\bf k}\!\uparrow, -{\bf k}\!\downarrow)$ excitations with $|\Delta_{\bf k}|>0$ is
\begin{widetext}
\begin{equation}
F_{s{\bf k}} = \xi_{\bf k}^{(0)} - E^{(s)}_{\bf k} + |\Delta_{\bf k}|^2\frac{\tanh\left(|\xi_{\bf
k}^{(0)}|/2\right)}{2|\xi_{\bf k}^{(0)}|} - T \ln \left[ 1+ e^{-\left(E^{(s)}_{\bf k}
 + \frac{\pi}{\sqrt{2}} az_{\bf k} \right)/T} \right] \left[ 1+ e^{-\left(E^{(s)}_{\bf k}
 - \frac{\pi}{\sqrt{2}} az_{\bf k} \right)/T} \right], \label{Fsk}
\end{equation}
where $E^{(s)}_{\bf k}=\sqrt{ \left( \xi^{(0)}_{\bf k} \right)^2 + |\Delta_{\bf k}|^2 }$.  For $|\Delta_{\bf k}|=0$,
it becomes
\begin{equation}
F_{n{\bf k}} = \xi_{\bf k}^{(0)} - |\xi_{\bf k}^{(0)}| - T \ln \left[ 1+ e^{-\left(|\xi_{\bf k}^{(0)}| +
\frac{\pi}{\sqrt{2}} az_{\bf k} \right) /T} \right]  \left[ 1+ e^{-\left(|\xi_{\bf k}^{(0)}| - \frac{\pi}{\sqrt{2}}
az_{\bf k} \right) /T} \right]. \label{Fnk}
\end{equation}
\end{widetext}
Both $F_{s{\bf k}}$ and $F_{n{\bf k}}$ are functions of $|\xi_{\bf k}^{(0)}|$, $T$ and $az_{\bf k}$. For given
$|\xi_{\bf k}^{(0)}|$, $T$ and $az_{\bf k}$, the $|\Delta_{\bf k}|>0$ solution is the stable solution if $F_{s{\bf
k}} < F_{n{\bf k}}$. Otherwise, the $|\Delta_{\bf k}|=0$ solution is the stable solution.

We define temperature $T_{c\bf k}$ and vector potential magnitude $a_{c\bf k}$ as such that the $|\Delta_{\bf k}|=0$
solution becomes the stable solution for $T\ge T_{c\bf k}$ or $a\ge a_{c\bf k}$.  By definition, we have $T_{c\bf
k}\leq T^{\star}_{c\bf k}$ and $a_{c{\bf k}}\leq a^{\star}_{c\bf k}$.  Note that a pair of $({\bf k}\!\uparrow,-{\bf
k}\!\downarrow)$ electrons becomes de-paired (having $|\Delta_{\bf k}|=0$) for $T\geq T_{c{\bf k}}$ or  $a\geq
a_{c\bf k}$.

\begin{figure}[t]
\includegraphics{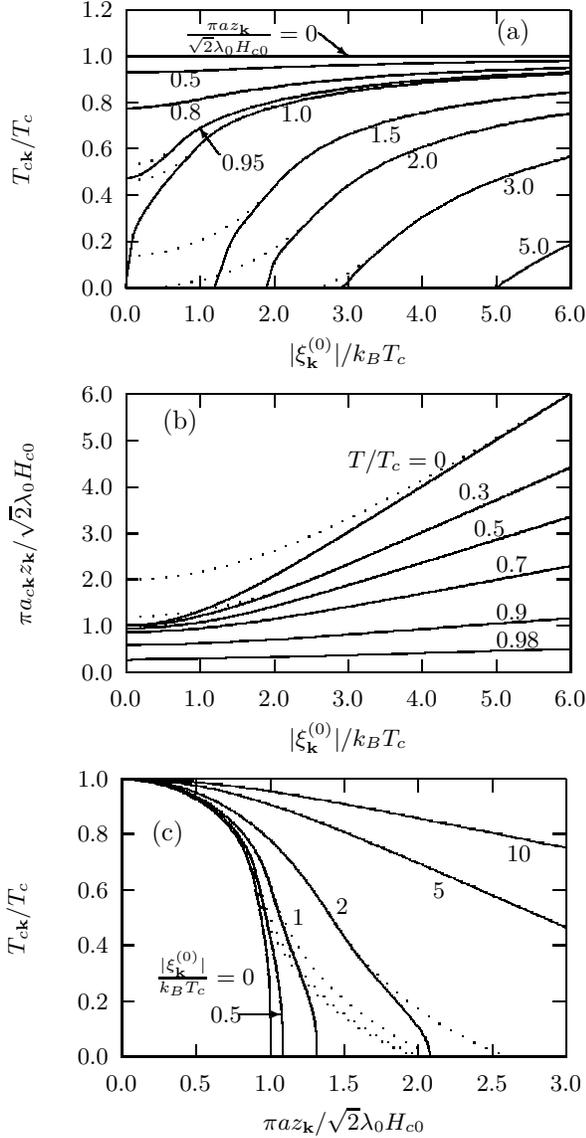}
\caption{(a) $T_{c{\bf k}}$ versus $|\xi_{\bf k}^{(0)}|$ for different values of $a z_{\bf k}$; (b) $a_{c\bf k}z_{\bf
k}$ versus $|\xi_{\bf k}^{(0)}|$ for different values of $T$; (c) $T_{c{\bf k}}$ versus $az_{\bf k}$ for different
values of $|\xi^{(0)}_{\bf k}|$. The dotted curves show the corresponding results for $T^{\star}_{c{\bf k}}$ or
$a^{\star}_{c{\bf k}}z_{\bf k}$.} \label{figure1}
\end{figure}

\begin{figure}[t]
\includegraphics{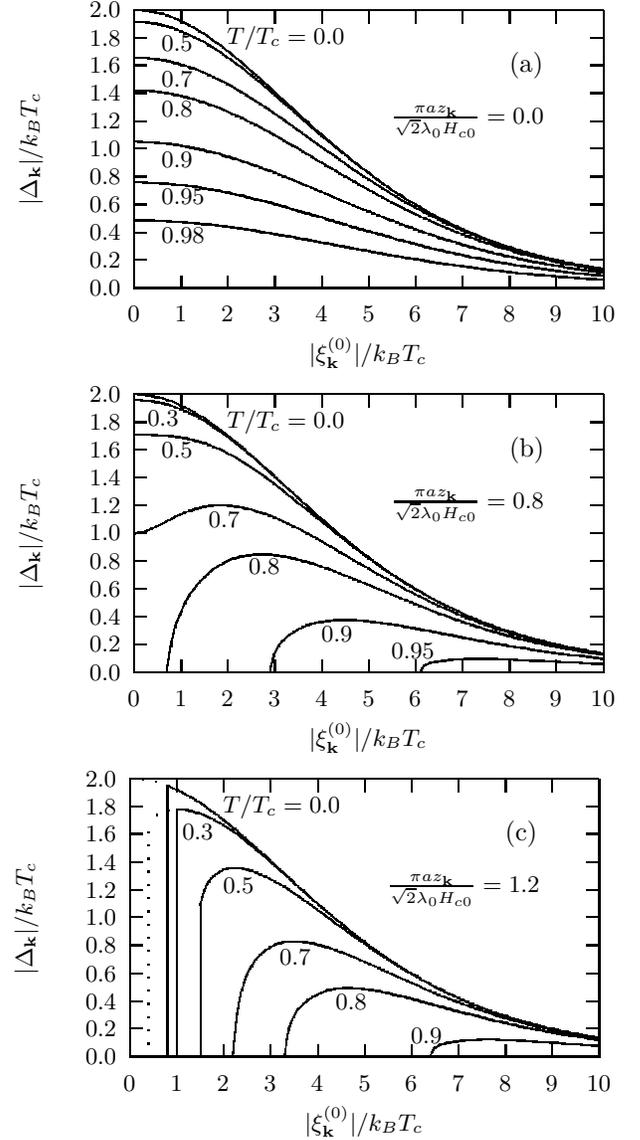}
\caption{ $|\Delta_{\bf k}|$ versus $|\xi_{\bf k}^{(0)}|$ for different values of $T$ and $az_{\bf k}$;\, $\pi
az_{\bf k}/\sqrt{2}\lambda_0H_{c0}=0$, $0.8$, and $1.2$ in (a), (b), and (c), respectively. The values of $T/T_c$ are
indicated on the curves. Unstable solutions are shown by dotted curves.} \label{figure2}
\end{figure}

\begin{figure}
\includegraphics{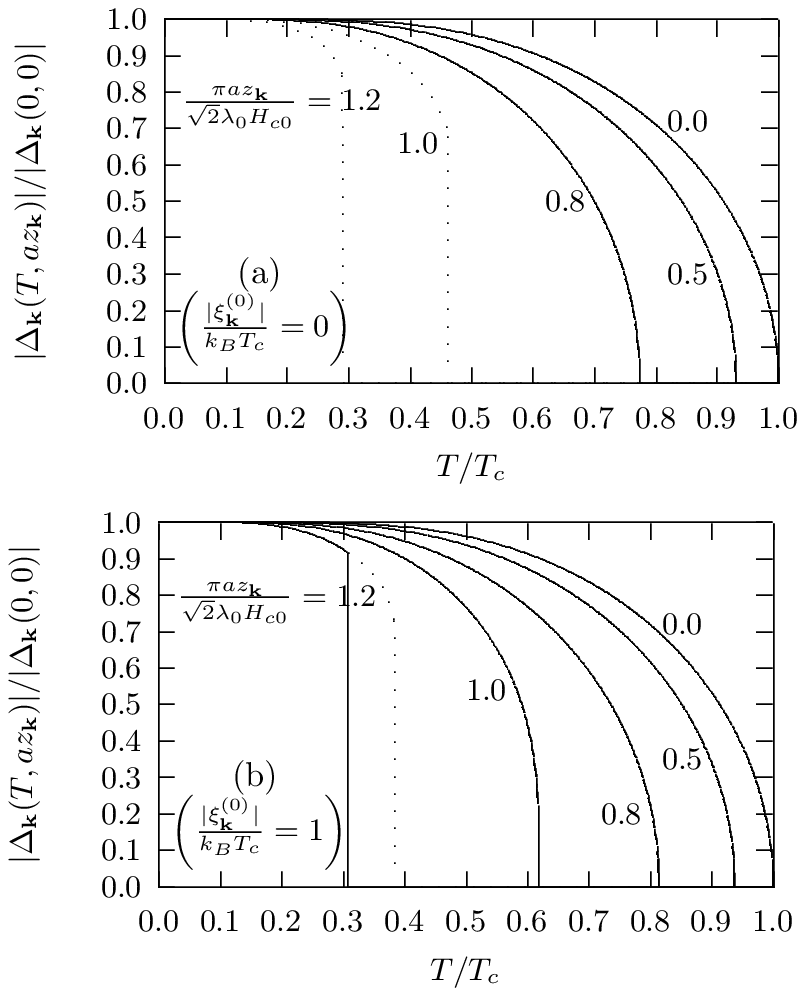}
\caption{$|\Delta_{\bf k}(T,az_{\bf k})|/|\Delta_{\bf k}(0,0)|$ versus $T$ for different values of $|\xi_{\bf
k}^{(0)}|$ and $az_{\bf k}$;\, $|\xi_{\bf k}^{(0)}|/k_BT_c=0$ and $1.0$ in (a) and (b), respectively. The values of
$\pi az_{\bf k}/\sqrt{2}\lambda_0H_{c0}$ are indicated on the curves. Unstable solutions are shown by the dotted
curves.} \label{figure3}
\end{figure}

\begin{figure}
\includegraphics{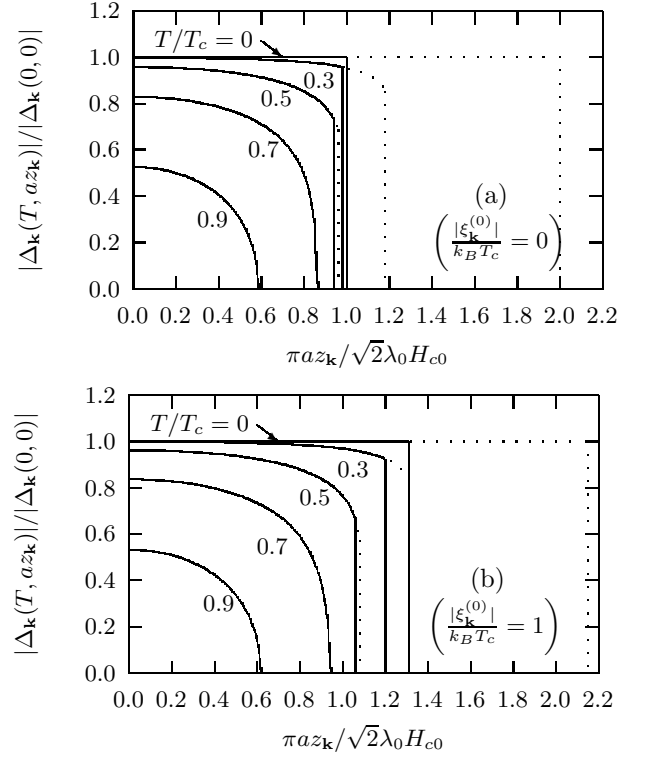}
\caption{ $|\Delta_{\bf k}(T,az_{\bf k})|/|\Delta_{\bf k}(0,0)|$ versus $az_{\bf k}$ for different values of
$|\xi_{\bf k}^{(0)}|$ and $T$;\, $|\xi_{\bf k}^{(0)}| /k_BT_c= 0$ and $1.0$ in (a) and (b), respectively. The values
of $T/T_c$ are indicated on the curves. Unstable solutions are shown by the dotted curves.} \label{figure4}
\end{figure}

\begin{figure}
\includegraphics{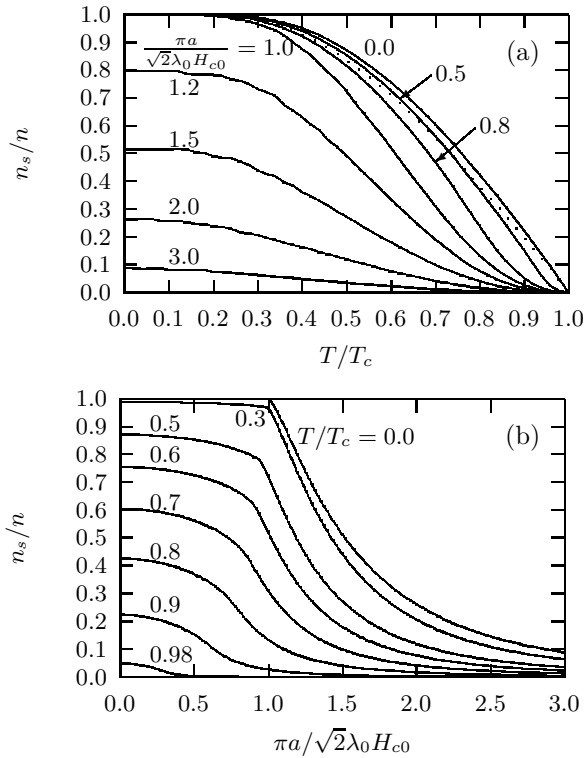}
\caption{(a) ``Effective superconducting electron density'' $n_s$ versus $T$ for different values of $a$ as indicated
on the curves; (b) $n_s$ versus $a$ for different values of $T$ as indicated on the curves. The dotted curve in (a)
shows M\"{u}hlschlegel's result\cite{mu} for $\lambda^2_0/\lambda^2_L(T)$, where $\lambda_L(T) = \lim_{H_a\rightarrow
0}\lambda(T,H_a)$. } \label{figure5}
\end{figure}

\begin{figure}
\includegraphics{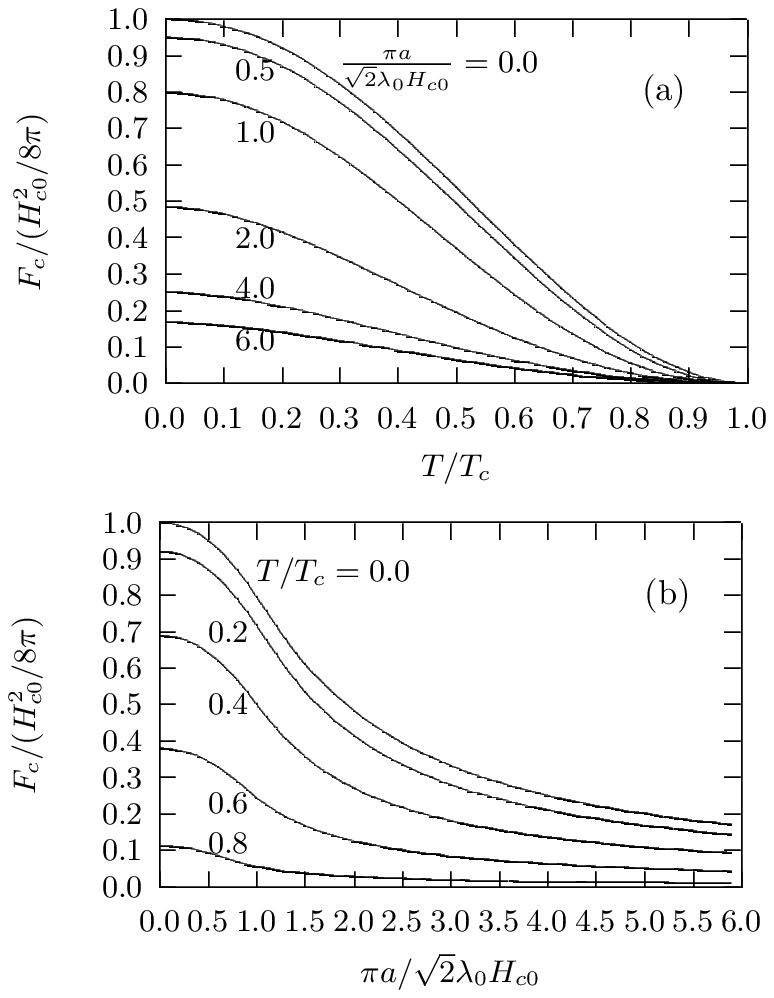}
\caption{(a) Condensation energy density $F_c$ versus $T$  for different values of $a$ as indicated on the curves;
(b) $F_c$ versus $a$ for different values of $T$ as indicated on the curves. } \label{figure6}
\end{figure}

Numerical results for $T_{c{\bf k}}$ and $a_{c{\bf k}}$ are shown in Fig. 1  [Figs. 1(a)-1(c)].  Figure 1(a) shows
$T_{c{\bf k}}$ versus $|\xi^{(0)}_{\bf k}|$ for different values of $a z_{\bf k}$; Fig. 1(b) shows $a_{c{\bf
k}}z_{\bf k}$ versus $|\xi^{(0)}_{\bf k}|$ for different values of $T$; and Fig. 1(c) shows $T_{c{\bf k}}$ versus
$az_{\bf k}$ for different values of $|\xi^{(0)}_{\bf k}|$. The dotted curves in Fig. 1 show corresponding results
for $T^{\star}_{c{\bf k}}$ and $a^{\star}_{c{\bf k}}z_{\bf k}$.

As shown in Figs. 1(a) and 1(b), both $T_{c{\bf k}}$ and $a_{c{\bf k}}z_{\bf k}$ are increasing functions of
$|\xi^{(0)}_{\bf k}|$, meaning that pairs of electrons with smaller $|\xi^{(0)}_{\bf k}|$ become de-paired at lower
values of $T$ and/or $az_{\bf k}$ than those with larger $|\xi^{(0)}_{\bf k}|$.  Figure 1(c) shows that $T_{c{\bf
k}}$ is a decreasing function of $az_{\bf k}$, which also means that $a_{c{\bf k}}z_{\bf k}$ is a decreasing function
of $T$.

For constant $|\xi^{(0)}_{\bf k}|$, the values of $T_{c{\bf k}}$ and $a_{c{\bf k}}$ are lowest when {\bf k} is
parallel to {\bf a}, because {\bf a} appears in Eq. (\ref{haoeq1}) only in the form of $az_{\bf k}$, which is largest
(therefore, most effective as a de-pairing force) when $z_{\bf k}=1$. For two electronic states with wave-vectors
${\bf k'}$ and {\bf k}, respectively, if $|\xi^{(0)}_{\bf k'}| =|\xi^{(0)}_{\bf k}|$ and {\bf k} is parallel to {\bf
a}, then, the following relations hold: $T_{c{\bf k'}}(a) = T_{c{\bf k}\|{\bf a}}(az_{\bf k'})$; and $a_{c{\bf
k'}}(T) = a_{c{\bf k}\|{\bf a}}(T)/z_{\bf k'}$.

Note that $T_{c{\bf k}_F\|{\bf a}}(a)$ is the lowest value of $T_{c{\bf k}}(a)$, i.e., $T_{c{\bf k}_F\|{\bf a}}(a) =
T_{c{\bf k},\text{min}}(a)$.  Similarly, $a_{c{\bf k}_F\|{\bf a}}(T)$ is the lowest value of $a_{c{\bf k}}(T)$, i.e.,
$a_{c{\bf k}_F\|{\bf a}}(T) = a_{c{\bf k},\text{min}}(T)$.  The function $T_{c{\bf k}_F\|{\bf a}}(a)$ [or its inverse
function $a_{c{\bf k}_F\|{\bf a}}(T)$], which is shown by the $|\xi^{(0)}_{\bf k}|=0$ curve in Fig. 1(c) (for $z_{\bf
k}=1$), is of particular importance.  When $T < T_{c{\bf k}_F\|{\bf a}}(a)$ [or $a < a_{c{\bf k}_F\|{\bf a}}(T)$] in
a region in a superconductor [remember that ${\bf a}={\bf a}({\bf x})$ is a location-dependent quantity], this region
is in a ``all-paired state,'' in which $|\Delta_{\bf k}|>0$ for all the electrons. On the other hand, when $T >
T_{c{\bf k}_F\|{\bf a}}(a)$ [or $a > a_{c{\bf k}_F\|{\bf a}}(T)$] in a region in a superconductor, this region is in
a ``partly-paired state,'' in which electrons with $T_{c{\bf k}}(az_{\bf k})$ in the range $T_{c{\bf k}_F\|{\bf
a}}(a)< T_{c{\bf k}}(az_{\bf k}) < T$ [or with $a_{c{\bf k}}(T)$ in the range $a_{c{\bf k}_F\|{\bf a}}(T) < a_{c{\bf
k}}(T) < a$] become de-paired (having $|\Delta_{\bf k}|=0$), while electrons with $T_{c{\bf k}}(az_{\bf k}) > T$ [or
$a_{c{\bf k}}(T)>a$] remain paired (having $|\Delta_{\bf k}| > 0$).

Numerical results for $|\Delta_{\bf k}|$ versus $|\xi^{(0)}_{\bf k}|$ for different values of $T$ and $az_{\bf k}$
are shown in Fig. 2 [Figs. 2(a)-2(c)]. Figure 2(a) shows the case of ${\bf a}=0$.  In this case, $|\Delta_{\bf k}|$
is a monotonic decreasing function of $|\xi^{(0)}_{\bf k}|$ for all temperatures below $T_c$; and $|\Delta_{\bf k}|$
vanishes at the same temperature $T_c$ for all values of $|\xi_{\bf k}^{(0)}|$.

Figure 2(b) shows an example of the case of $0<a<a_{c{\bf k}_F\|{\bf a}}(0)$ [note that $\pi a_{c{\bf k}_F\|{\bf
a}}(0)/\sqrt{2}=1$ (in dimensionless units)]. In this case, the $|\Delta_{\bf k}|$-versus-$|\xi^{(0)}_{\bf k}|$ curve
for $T=0$ is the same as in the case of ${\bf a}=0$. However, as $T$ increases, $|\Delta_{\bf k}|$ for a smaller
$|\xi^{(0)}_{\bf k}|$ is more strongly suppressed, and decreases faster, so that the $|\Delta_{\bf
k}|$-versus-$|\xi^{(0)}_{\bf k}|$ curve eventually becomes non-monotonic, with a maximum located away from $|\xi_{\bf
k}^{(0)}|=0$. As $T$ increases further, $|\Delta_{\bf k}|$ for a smaller $|\xi^{(0)}_{\bf k}|$ vanishes at a lower
temperature [namely, $T_{c{\bf k}}$ is smaller for smaller $|\xi_{\bf k}^{(0)}|$, a feature that is also shown by
Fig. 1(a)]. The $|\Delta_{\bf k}|$-versus-$|\xi^{(0)}_{\bf k}|$ curve then has two parts: a $|\Delta_{\bf k}|=0$ part
for low energies, for which $T_{c{\bf k}}(az_{\bf k})<T$ [or $a_{c{\bf k}}(T)<a$], and a $|\Delta_{\bf k}|>0$ part
for higher energies, for which $T_{c{\bf k}}(az_{\bf k})>T$ [or $a_{c{\bf k}}(T)>a$].

Figure 2(c) shows an example of the case of $a>a_{c{\bf k}_F\|{\bf a}}(0)$. In this case, the $|\Delta_{\bf
k}|$-versus-$|\xi^{(0)}_{\bf k}|$ curve has a $|\Delta_{\bf k}|=0$ part even for $T=0$. Namely, at $T=0$,
$|\Delta_{\bf k}|=0$ for those electronic states with $a_{c{\bf k}}(0)<a$.  The vertical rises in the $|\Delta_{\bf
k}|$-versus-$|\xi^{(0)}_{\bf k}|$ curves for low temperatures (i.e., the curves for $T=0;$ 0.3 and 0.5) in Fig. 2(c)
indicate discontinuities. The doted curves in Fig. 2(c) show corresponding unstable solutions of Eq. (\ref{haoeq1})
for $T$ in the range $T_{c{\bf k}}<T<T_{c{\bf k}}^{\star}$ (or for $a$ in the range $a_{c{\bf k}}<a<a_{c{\bf
k}}^{\star}$).

Figures 3 and 4 show, respectively, the $T$-dependence and $az_{\bf k}$-dependence of $|\Delta_{\bf k}|$.  The dotted
curves in Figs 3 and 4 show corresponding unstable solutions of Eq. (\ref{haoeq1}) in the range $T_{c{\bf
k}}<T<T_{c{\bf k}}^{\star}$ or $a_{c{\bf k}}<a<a_{c{\bf k}}^{\star}$. As shown in the figures, $|\Delta_{\bf k}|$ is
a monotonic decreasing function of $T$ and $az_{\bf k}$, except that at $T=0$, $|\Delta_{\bf k}|>0$ is a constant for
$a<a_{c{\bf k}}(0)$.  Note that, at $T=T_{c{\bf k}}$ or $a = a_{c{\bf k}}$, $|\Delta_{\bf k}|$ may become zero
continuously or discontinuously, depending on $|\xi^{(0)}_{\bf k}|$, $T$ and $az_{\bf k}$.  The vertical drops in
some of the curves shown in Figs. 3 and 4 indicate discontinuities.

Based on numerical solutions for $|\Delta_{\bf k}(T,{\bf a})|$, we can compute $n_s(T,a)$ and $F_c(T,a)$
straightforwardly by numerically carrying out the integrals involved in the expressions for $n_s$ and $F_c$.
Numerical results for $n_s(T,a)$ and $F_c(T,a)$ are shown in Figs. 5 and 6, respectively.

The dotted curve in Fig. 5(a) shows M\"{u}hlschlegel's numerical result\cite{mu} for $\lambda^2_0/\lambda^2_L(T)$,
which is based on a linear-response approach,\cite{bcs} and therefore, is valid only in the low-field limit, i.e.,
$\lambda_L(T)=\lim_{H_a\rightarrow 0}\lambda(T,H_a)$.  [More discussion on magnetic field penetration depth
$\lambda(T,H_a)$ is given below in Sec. \ref{secLambda}.]  Note that $\lambda^2_0/\lambda^2_L(T)$ corresponds to our
$n_s(T,a)/n$ for $a = 0$. The quantitative difference between M\"{u}hlschlegel's $\lambda^2_0/\lambda^2_L(T)$ and our
$n_s(T,0)/n$, as shown in Fig. 5(a), is due to the use of the cut-off approximation\cite{bcs} for solving the energy
gap equation in M\"{u}hlschlegel's work.

The $F_c$-versus-$T$ curve for $a=0$ in Fig. 6(a) shows $F_c(T,0)$, which is the same as $H_c^2(T)/8\pi$.  This
result was also previously obtained by the author in Ref. \onlinecite{hao96}.

As shown in Figs. 5 and 6, $n_s(T,a)$ and $F_c(T,a)$ both are monotonic decreasing functions of $T$ and $a$, except
that $n_s\!=1$ is a constant for $a \leq a_{c{\bf k}_F\|{\bf a}}(0)$ (where $\pi a_{c{\bf k}_F\|{\bf a}}(0)/\sqrt{2}
= 1$) at $T=0$ [Fig. 5(b)]. Note that, for $T=0$ and $a \leq a_{c{\bf k}_F\|{\bf a}}(0)$, we have $F_c\!= 1-a^2$
[which can be derived either from the expression for $F_c$, or by solving Eq. (\ref{nsFc}) for $n_s=1$ and
$F_c(0,0)=1$]. This is shown in Fig. 6(b), where we can see that the $F_c$-versus-$a$ curve for $T=0$ is parabolic
for $0\le \pi a/\sqrt{2} \le 1$. Namely, in this case, all the electrons are paired and no thermal excitations exist,
so that the decrease in $F_c$ (i.e., the $-a^2$ term) is entirely due to the kinetic energy associated with the
uniform flow of all the electrons. At $T=0$, as $a$ increases further so that $a > a_{c{\bf k}_F\|{\bf a}}(0)$ (i.e.,
$\pi a/\sqrt{2}\!~>~\!1$), de-paired electrons (having $|\Delta_{\bf k}|=0$) begin to appear, and we have $n_s<1$.

The present analysis provides a possible explanation for the experimentally observed non-vanishing Knight shifts in
the superconducting state near $T=0$ (for example, Refs. \onlinecite{reif} and \onlinecite{androes}). Namely, at
$T=0$, electrons in a superconducting sample are all paired only when the applied magnetic field is weak so that we
have $a < a_{c{\bf k}_F\|{\bf a}}(0)$ everywhere in the sample; this gives zero spin-polarization, and therefore,
zero Knight shift (which is proportional to the density of spin polarization\cite{slichter}). However, we note that
the applied magnetic fields that were used for the Knight shift measurements\cite{reif,androes} are comparable to the
thermodynamic critical magnetic fields of the samples. Therefore, it is likely that we actually had $a > a_{c{\bf
k}_F\|{\bf a}}(0)$ over a significant portion of the sample, where a finite fraction of the electrons were de-paired
and spin-polarized, giving rise to a non-zero Knight shift. The previous theoretical prediction of a zero Knight
shift in the superconducting state for $T\rightarrow 0$ by Yosida\cite{yosida} is valid only for weak magnetic
fields. Although the present theory is developed for highly-local superconductors, we expect qualitative conclusions
of the theory, including the prediction for a non-vanishing Knight shift near $T=0$ for a not-so-weak magnetic field,
to be valid also for non-local superconductors.

\section{ Applications }
\label{secAppl}

We present in this section a few examples of application of the theory.   We consider the case of a semi-infinite
superconductor in an applied magnetic field ${\bf H}_a$ parallel to the surface of the superconductor and the case of
an isolated vortex in an infinite superconductor.  We determine, in each case, spatial variations of vector potential
{\bf a}, magnetic flux density {\bf b}, electrical current density {\bf j}, energy gap parameter amplitude
$|\Delta_{\bf k}|$, ``effective superconducting electron density'' $n_s$ and condensation energy density $F_c$. We
also calculate magnetic field penetration depth $\lambda(T,H_a)$ and lower critical magnetic field $H_{c1}(T)$.

\subsection{ Semi-infinite superconductor }
\label{secSemiInfiniteSC}

We consider a semi-infinite superconductor in an applied magnetic field ${\bf H}_a$ parallel to the surface of the
superconductor. Let the superconductor occupy the half space $x\!>\!0$ and ${\bf H}_a$ be applied along the $z$-axis.
In terms of the Cartesian coordinates $(x,y,z)$ and the unit vectors $(\hat{\bf x}, \hat{\bf y}, \hat{\bf z})$, we
can write ${\bf b} = b(x)\hat{\bf z}$, ${\bf j} = j(x)\hat{\bf y}$, and ${\bf a} = - a(x)\hat{\bf y}$. Then,
Eq.~(\ref{haoeq2}) and the relation ${\bf b}=\nabla\!\times\!{\bf a}$ become
\begin{equation}
b'(x) = -n_s(x)a(x) \label{bprime}
\end{equation}
and
\begin{equation}
a'(x) = -b(x)\,, \label{aprime}
\end{equation}
respectively, where a ``prime'' indicates a derivative with respect to $x$; $n_s(x)=n_s(a(x))$ is given by Eq.
(\ref{ns-eq}), where $|\Delta_{\bf k}(x)|=|\Delta_{\bf k}(a(x))|$ is determined by Eq. (\ref{haoeq1}); and the
relation $j(x)=-b'(x)$ has been used.

This is a non-linear second-order boundary-value problem with boundary conditions
\begin{equation}
b(0)=-a'(0)=H_a \label{bcAt0}
\end{equation}
and
\begin{equation}
a(\infty)=b(\infty)=0. \label{bcAtInfty}
\end{equation}

Note that, for the convenience of numerical calculation, we have expressed this second-order boundary-value problem
as a system of two first-order differential equations. The numerical method for solving this boundary-value problem
is explained in Appendix \ref{computeAB}.

Numerical results for $a(x)$, $b(x)$, $j(x)$, $n_s(x)$, $|\Delta_{{\bf k}_F\|{\bf a}}(x)|$ and $F_c(x)$ near the
surface of the superconductor for different values of $H_a$ are shown in Fig. 7 [Figs. 7(a)-7(c)] and Fig. 8 [Figs.
8(a)-8(c)] for $T=0$ and 0.6, respectively. As an example of $|\Delta_{\bf k}(x)|$, $|\Delta_{{\bf k}_F\|{\bf
a}}(x)|$ is shown in the figures. Here $\Delta_{{\bf k}_F\|{\bf a}}$ is the energy gap parameter of an electronic
state on the Fermi surface with wave vector ${\bf k}_F$ parallel to {\bf a}, of which the value of $T_{c{\bf k}}(a)$,
or $a_{c{\bf k}}(T)$, is the lowest among all the electronic states, i.e., $T_{c{\bf k}_F\|{\bf a}}(a)=T_{c{\bf
k},\text{min}}(a)$, and $a_{c{\bf k}_F\|{\bf a}}(T)=a_{c{\bf k},\text{min}}(T)$, as we mentioned earlier.

Figures 7(a) and 8(a) show examples of the case where $H_a$ is low so that $a<a_{c{\bf k}_F\|{\bf a}}(T)$ at the
surface of the superconductor. In this case, no de-paired electrons exist in the superconductor, i.e., $|\Delta_{\bf
k}|>0$ for all the electrons in the superconductor. At $T=0$ [Fig. 7(a)], $|\Delta_{\bf k}|$ and $n_s$ are
$x$-independent [because $|\Delta_{\bf k}|$ and $n_s$ are $a$-independent for $a<a_{c{\bf k}_F\|{\bf a}}$ at $T=0$,
as shown in Figs. 4 and 5, respectively], and therefore, our results for $a(x)$, $b(x)$, $j(x)$ and $F_c(x)$ are the
same as those of the London theory. Namely, in this case, our Eq. (\ref{haoeq2}) is the same as the London equation,
which gives linear relations between $a(x)$, $b(x)$ and $j(x)$, i.e., $j(x)\propto b(x)\propto a(x)$.  For $T>0$
[Fig. 8(a)], $|\Delta_{\bf k}|$ and $n_s$ become $x$-dependent near the surface of the superconductor, and therefore,
our results for $a(x)$, $b(x)$, $j(x)$ and $F_c(x)$ near the surface of the superconductor are no longer the same as
those of the London theory.

Figures 7(b), 7(c), 8(b) and 8(c) show examples of the case where $H_a$ has been increased to a point such that
$a>a_{c{\bf k}_F\|{\bf a}}(T)$ at the surface of the sample.  Let $x_0$ be the coordinate of the location at which
$a=a_{c{\bf k}_F\|{\bf a}}(T)$.  We have $|\Delta_{{\bf k}_F\|{\bf a}}(x)| = 0$ in the region $0\le x < x_0$, and
$|\Delta_{{\bf k}_F\|{\bf a}}(x)| >0$ in the region $x>x_0$. The region $0\le x < x_0$ is in a ``partly-paired
state,'' where paired electrons (having $|\Delta_{\bf k}|> 0$) and de-paired electrons (having $|\Delta_{\bf k}| =
0$) co-exist. The region $x>x_0$ remains ``all-paired,'' where $|\Delta_{\bf k}|> 0$ for all the electrons. Our
results for $a(x)$, $b(x)$, $j(x)$ and $F_c(x)$ in the region $0\le x < x_0$ are significantly different from those
of the London theory, especially when $H_a$ is high [Figs. 7(c) and 8(c)]. In particular, $j(x)$ is no longer a
monotonic decreasing function of $x$, having a maximum located near $x_0$.

\begin{figure}
\includegraphics{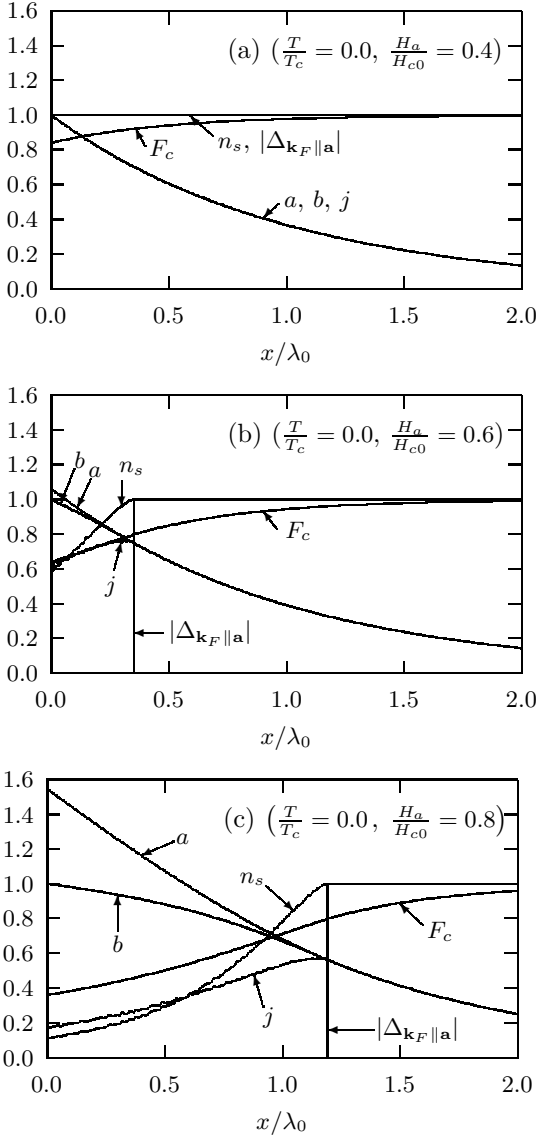}
\caption{Variations of $a(x)$, $b(x)$, $j(x)$, $n_s(x)$, $|\Delta_{{\bf k}_F\|{\bf a}}(x)|$, and $F_c(x)$ near the
surface at $T=0$ and different values of $H_a$: $H_a/H_{c0}=0.4$ (a), $0.6$ (b), and $0.8$ (c). The quantities
$a(x)$, $b(x)$, $j(x)$, $n_s(x)$, $|\Delta_{{\bf k}_F\|{\bf a}}(x)|$, and $F_c(x)$ are measured in units of
$\lambda_0 H_a$, $H_a$, $H_a/\lambda_0$, $n_s(\infty)$, $|\Delta_{{\bf k}_F}(\infty)|$, and $F_c(\infty)$,
respectively. } \label{figure7}
\end{figure}

\begin{figure}
\includegraphics{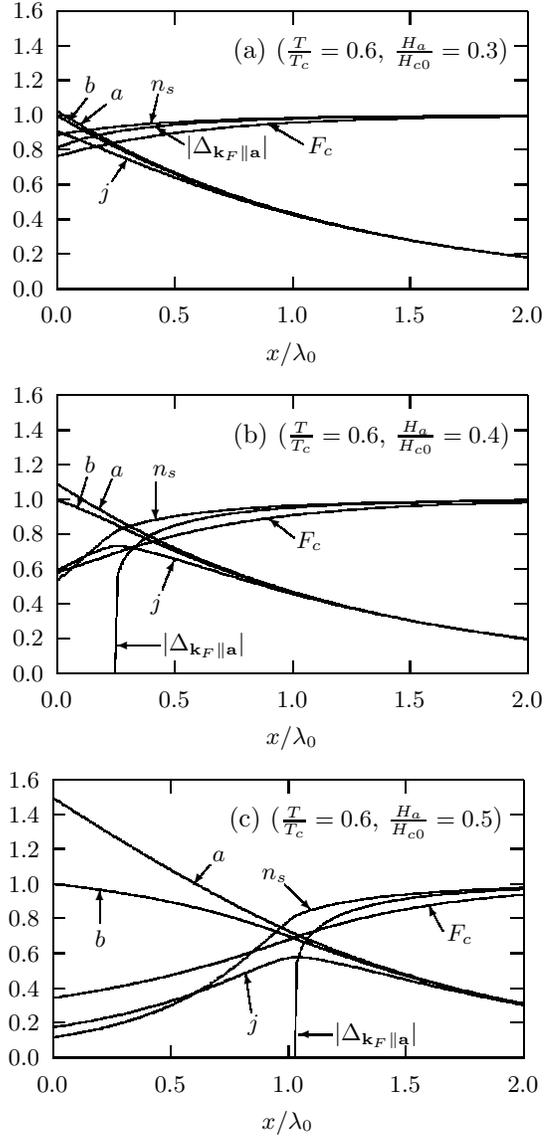}
\caption{Variations of $a(x)$, $b(x)$, $j(x)$, $n_s(x)$, $|\Delta_{{\bf k}_F\|{\bf a}}(x)|$, and $F_c(x)$ near the
surface at $T/T_c=0.6$ and different values of $H_a$: $H_a/H_{c0}=0.3$ (a), $0.4$ (b), and $0.5$ (c)
[$H_a/H_c(T)=0.49$ (a), $0.65$ (b), and $0.81$ (c)]. The quantities $a(x)$, $b(x)$, $j(x)$, $n_s(x)$, $|\Delta_{{\bf
k}_F\|{\bf a}}(x)|$, and $F_c(x)$ are measured in units of $\lambda_L(T) H_a$, $H_a$, $H_a/\lambda_L(T)$,
$n_s(\infty)$, $|\Delta_{{\bf k}_F}(\infty)|$, and $F_c(\infty)$, respectively, where $\lambda_L(T) =
\lim_{H_a\rightarrow 0} \lambda(T,H_a)$.} \label{figure8}
\end{figure}

\begin{figure}
\includegraphics{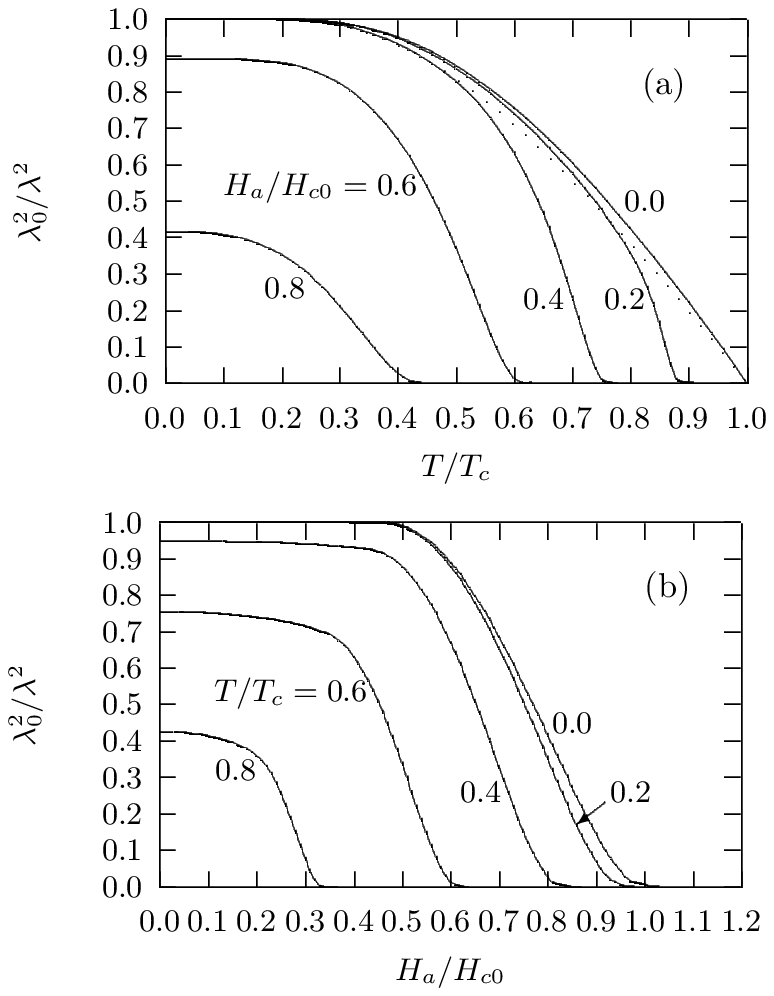}
\caption{(a) $\lambda_0^2/\lambda^2$ versus $T$ for different values of $H_a$ as indicated on the curves; (b)
$\lambda_0^2/\lambda^2$ versus $H_a$ for different values of $T$ as indicated on the curves. The dotted curve in (a)
shows the M\"{u}hlschlegel's result\cite{mu} for $\lambda^2_0/\lambda_L^2(T)$.} \label{figure9}
\end{figure}

\subsection{ Magnetic field penetration depth $\lambda(T,H_a)$ }
\label{secLambda}

The solution for $a(x)$ obtained in Sec. \ref{secSemiInfiniteSC} for the case of a semi-infinite superconductor can
be used to calculate magnetic field penetration depth $\lambda(T,H_a)$, which, for a semi-infinite superconductor, is
defined as\cite{GL}
\begin{equation}
\label{lambda-eq} \lambda = \frac{1}{H_a} \int_0^\infty \!\!\! b(x) dx = \frac{a(0)}{H_a},
\end{equation}
where $a(0)$ is the value of $a(x)$ at the surface of the superconductor.

Numerical results for $\lambda(T,H_a)$ are shown as $\lambda^2_0/\lambda^2$ versus $T$ for different values of $H_a$
in Fig. 9(a), and $\lambda^2_0/\lambda^2$ versus $H_a$ for different values of $T$ in Fig. 9(b).

The dotted curve in Fig. 9(a) shows M\"{u}hlschlegel's numerical result\cite{mu} for $\lambda^2_0/\lambda^2_L(T)$. As
we mentioned earlier, the M\"{u}hlschlegel's result is based on a linear-response approach, and corresponds to our
$n_s(T,a)/n$ for $a=0$, which is the same as $\lambda^2_0/\lambda^2(T,H_a)$ for $H_a=0$, i.e.,
$\lambda^2_0/\lambda^2_L(T)=\lim_{a\rightarrow 0}n_s(T,a)/n=\lim_{H_a\rightarrow 0}\lambda^2_0/\lambda^2(T,H_a)$. The
quantitative difference between M\"{u}hlschlegel's $\lambda^2_0/\lambda^2_L(T)$ and our $n_s(T,a)/n$ for $a=0$, or
$\lambda^2_0/\lambda^2(T,H_a)$ for $H_a=0$, is due to the use of the cut-off approximation\cite{bcs} in
M\"{u}hlschlegel's work.

As shown in Fig. 9(b), for $H_a$ up to about $H_c(T)/2$, $\lambda_0^2/\lambda^2$ is $H_a$-independent or nearly
$H_a$-independent for low temperatures, and only weakly $H_a$-dependent for higher temperatures. However, for $H_a$
above about $H_c(T)/2$, $\lambda_0^2/\lambda^2$ drops rapidly as $H_a$ increases.  Note that the case of $H_a$ below
about $H_c(T)/2$ corresponds to the case shown in Figs. 7(a) and 8(a), where $a<a_{c{\bf k}_F\|{\bf a}}(T)$ at the
surface of the superconductor so that no de-paired electrons exist in the superconductor, whereas the case of $H_a$
above about $H_c(T)/2$ corresponds to the case shown in Figs. 7(b), 7(c), 8(b) and 8(c), where $a>a_{c{\bf k}_F\|{\bf
a}}(T)$ at the surface of the superconductor so that de-paired electrons exist in the region near the surface of the
superconductor.

As shown in Figs. 9(a) and 9(b), $\lambda_0^2/\lambda^2\rightarrow 0$ when $H_a\rightarrow H_c(T)$, or $T\rightarrow
T_c(H_a)$ [here $T_c(H_a)$ is the inverse function of $H_c(T)$]. However, we should note that this is not true for a
type-II superconductor, for which superconductivity is completely suppressed only at the upper critical magnetic
field $H_{c2}$, which is usually much higher than $H_c$ for a highly-local superconductor. This is because the
definition for $\lambda$, i.e., Eq. (\ref{lambda-eq}), is valid only when the superconductor is in the Meissner
state. In the mixed state of a type-II superconductor, in which vortices exist, this definition for $\lambda$ is no
longer valid. Further, we note that the definition of Eq. (\ref{lambda-eq}) is valid only for bulk samples with
dimensions much larger than the magnetic field penetration depth.

\subsection{ Isolated vortex }
\label{secSingleVortex}

We consider an isolated vortex in an infinite superconductor.  Let the vortex be centered on the $z$-axis. In terms
of the cylindrical coordinates $(r,\phi,z)$ and the unit vectors $(\hat{\bf r},\hat{\bf \phi},\hat{\bf z})$, we can
write ${\bf b}=b(r)\hat{\bf z}$, ${\bf j}=j(r)\hat{\bf \phi}$, and ${\bf a}=-a(r)\hat{\bf \phi}$. Then, Eq.
(\ref{haoeq2}) and the relation ${\bf b}=\nabla\!\times\!{\bf a}$ become
\begin{equation}
b'(r) = -n_s(r)a(r) \label{bprime2}
\end{equation}
and
\begin{equation}
a'(r) = -b(r) - \frac{a(r)}{r} \,, \label{aprime2}
\end{equation}
respectively, where a ``prime'' indicates a derivative with respect to $r$; $n_s(r)=n_s(a(r))$ is given by Eq.
(\ref{ns-eq}), where $|\Delta_{\bf k}(r)|=|\Delta_{\bf k}(a(r))|$ is determined by Eq. (\ref{haoeq1}); and the
relation $j(r)=-b'(r)$ has been used.

This is a non-linear second-order boundary-value problem with boundary conditions
\begin{equation}
b(\infty)=a(\infty)=0  \label{bcAtInfty2}
\end{equation}
and
\begin{equation}
\frac{1}{\kappa} = \int_0^{\infty}drr b(r) .  \label{bcFluxQuantum}
\end{equation}
The last boundary condition comes from flux quantization, i.e., $\Phi_0 = 2\pi\int_0^{\infty}drr b(r)$ in
conventional units, where $\Phi_0$ is the flux quantum.  The parameter $\kappa$ is defined as
\begin{equation}
\kappa=2\pi\lambda_0^2H_{c0}/\Phi_0 .  \label{kappa}
\end{equation}

Note the difference between the present definition for $\kappa$ and the one in the Ginzburg-landau (GL)
theory:\cite{GL} $\kappa^{(GL)}~=~2\sqrt{2}\pi\lambda^2H_c/\Phi_0$. Besides the difference between
$\lambda_0^2H_{c0}$ for $\kappa$ of Eq. (\ref{kappa}) and $\lambda^2H_c$ for $\kappa^{(GL)}$, there is an extra
factor $\sqrt{2}$ in the expression for $\kappa^{(GL)}$. In the GL theory,\cite{GL} $\kappa^{(GL)}$ can also be
expressed as a ratio between the magnetic field penetration depth and the coherence length. A similar expression for
$\kappa$ does not exist in the present theory, because the coherence effect (or the non-local effect) is not
accounted for in the present theory.

For the convenience of numerical calculation, the above-described second-order boundary-value problem has been
expressed as a system of two first-order differential equations.  The numerical method for solving this
boundary-value problem is explained in Appendix \ref{computeVortex}.

Numerical results for $a(r)$, $b(r)$, $j(r)$, $n_s(r)$, $|\Delta_{{\bf k}_F\|{\bf a}}(r)|$ and $F_c(r)$ near the
vortex core are shown in Fig. 10 for $\kappa = 5$ at several different temperatures as indicated in the figure.
Here, as an example of $|\Delta_{\bf k}(r)|$, $|\Delta_{{\bf k}_F\|{\bf a}}(r)|$ is shown in the figure.

Let $r_0$ denote the location at which $a=a_{c{\bf k}_F\|{\bf a}}(T)$.  We have $|\Delta_{{\bf k}_F\|{\bf a}}(r)|=0$
for $0\leq r < r_0$, and $|\Delta_{{\bf k}_F\|{\bf a}}(r)|>0$ for $r > r_0$.  Paired electrons (with $|\Delta_{\bf
k}|>0$) and de-paired electrons (with $|\Delta_{\bf k}|=0$) co-exist in the region $0\leq r < r_0$, whereas no
de-paired electrons exist (i.e., $|\Delta_{\bf k}|>0$ for all the electrons) in the region $r>r_0$.

\begin{figure}
\includegraphics{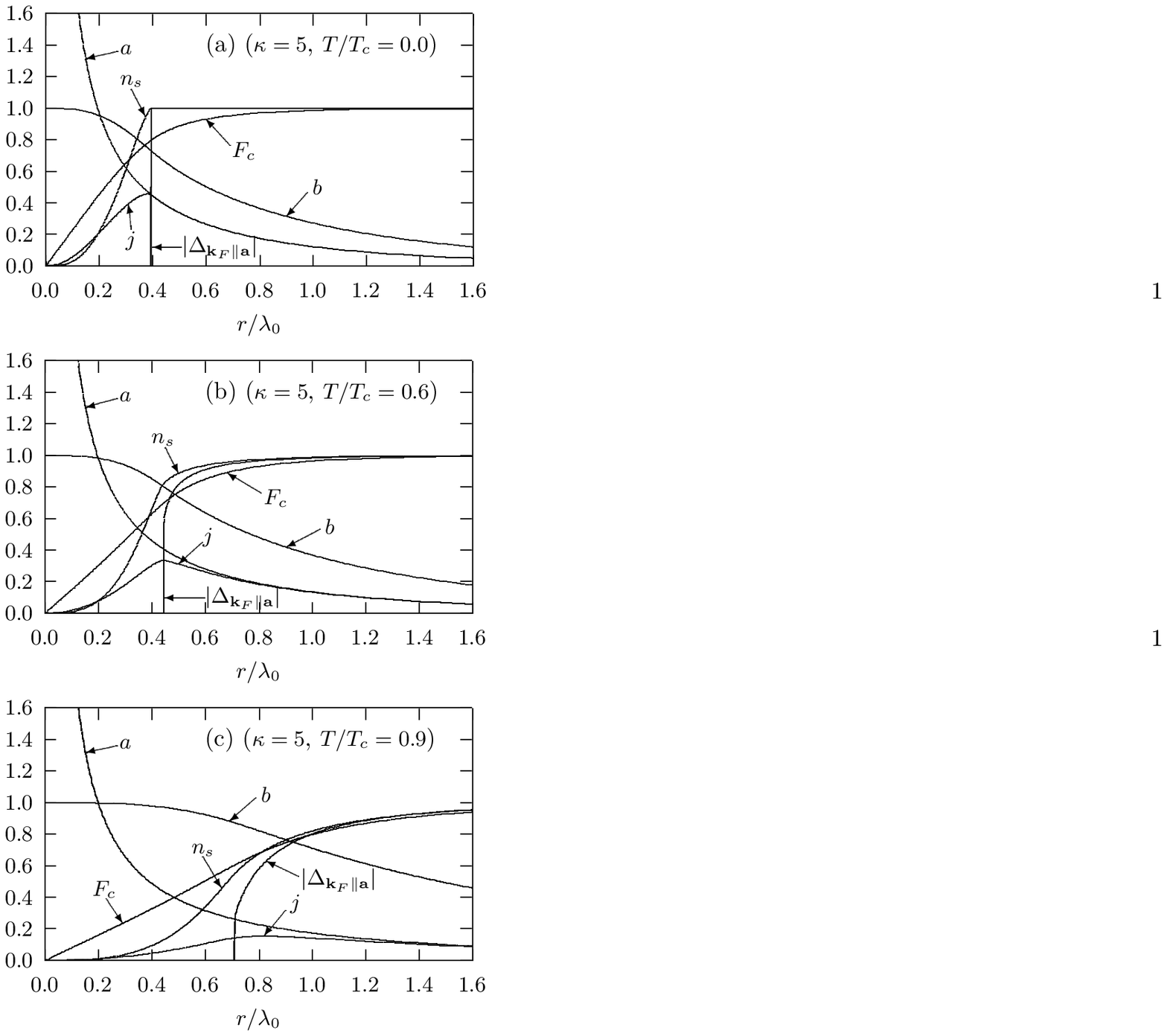}
\caption{Variations of $a(r)$, $b(r)$, $j(r)$, $n_s(r)$, $|\Delta_{{\bf k}_F\|{\bf a}}(r)|$, and $F_c(r)$ near the
vortex core at $T/T_c=0.0$ (a), $0.6$ (b), and $0.9$ (c), for $\kappa = 5$. The quantities $a(r)$, $b(r)$, $j(r)$,
$n_s(r)$, $|\Delta_{{\bf k}_F\|{\bf a}}(r)|$, and $F_c(r)$ are measured in units of $\lambda_0 H_{c0}$, $b(0)$,
$cH_{c0}/4\pi\lambda_0n_s(\infty)$, $n_s(\infty)$, $|\Delta_{{\bf k}_F}(\infty)|$, and $F_c(\infty)$, respectively. }
\label{figure10}
\end{figure}

\begin{figure}
\includegraphics{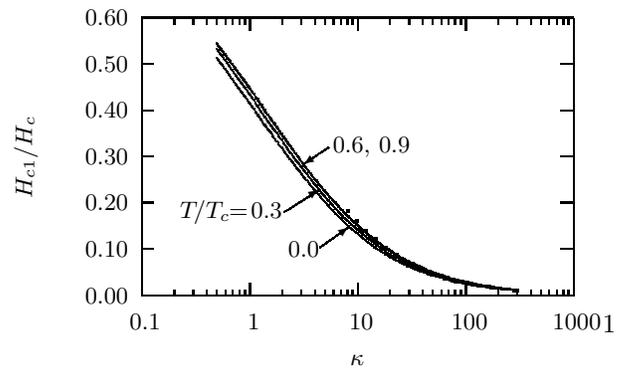}
\caption{ $H_{c1}/H_c$ versus $\kappa$ at different values of $T/T_c$ as indicated on the curves. The two curves for
$T/T_c=0.6$ and $0.9$ are practically indistinguishable from each other. The dotted curve shows the Ginzburg-Landau
results for $H_{c1}/H_c$ in the high-$\kappa$ limit. } \label{figure11}
\end{figure}

As shown in Fig. 10, as $r\rightarrow 0$, we have $a\rightarrow\infty$, $n_s\rightarrow 0$, $j\rightarrow 0$ and
$F_c\rightarrow 0$.  It also appears that $b(r)$, $j(r)$ and $n_s(r)$ have zero slopes at $r=0$, while $F_c(r)$ has a
finite slope at $r=0$ and is nearly linear for small $r$.  The electrical current density $j(r)$ has a maximum
located near $r_0$.

\subsection{ Lower critical magnetic field $H_{c1}(T)$ }

The numerical solutions obtained in the last subsection for an isolated vortex can be used to calculate lower
critical magnetic field $H_{c1}$ of the superconductor.  By definition,\cite{parksType2} at $H_a=H_{c1}$, the Gibbs
free energy must be the same whether the first vortex exists or not, i.e., ${\cal G}_{s}^{(\text{no vortex})} = {\cal
G}_{s}^{(\text{one vortex})}$ at $H_a=H_{c1}$. This condition leads to
\begin{equation}
H_{c1}=\frac{\kappa}{2}\int_0^{\infty}\!\!dr\,r\left(H_c^2-F_c+b^2\right).
\end{equation}

Numerical results for $H_{c1}$ are shown in Fig. 11 as $H_{c1}/H_c$ versus $\kappa$ for several different
temperatures.

The dotted curve in Fig. 11 shows the Ginzburg-Landau result for $H_{c1}/H_c$ for high-$\kappa$
superconductors:\cite{hu}
\begin{eqnarray}
\frac{H_{c1}^{(GL)}}{H_c} & = & \frac{1}{\sqrt{2}\kappa^{(GL)}} \left(\ln\kappa^{(GL)}\!+\!0.50\right) \nonumber \\
& = & \frac{1}{2\kappa}\left(\ln\kappa\!+~ \!0.85\right),
\end{eqnarray}
where we have used $\kappa^{(GL)}=\sqrt{2} \kappa$ by replacing $\lambda^2 H_c$ with $\lambda_0^2H_{c0}$ in the
expression for $\kappa^{(GL)}$.

As shown in Fig.  11, our result for $H_{c1}/H_c$ is only weakly $T$-dependent for low temperatures, and nearly
$T$-independent for intermediate and high temperatures, as indicated by the fact that the two
$H_{c1}/H_c$-versus-$\kappa$ curves for $T=0.6$ and $0.9$ are practically indistinguishable from each other. This
feature is to be compared with that $H_{c1}^{(GL)}/H_c$ is $T$-independent in the Ginzburg-Landau theory.  Figure 11
also shows that our result for $H_{c1}/H_c$ and the Ginzburg-Landau result are quantitatively not very different for
$\kappa\gg 1$.

For lower values of $\kappa$, the present theory underestimates the value of $H_{c1}$, because the coherence (or
non-local) effect, which increases the energy associated with a vortex and thus leads to a larger $H_{c1}$, is not
accounted for in the present theory.

\section{ Summary }
\label{secSummary}

We have presented a microscopic theory for superconductivity in a magnetic field based on a local approximation
approach.  The theory allows microscopic description of the suppression of superconductivity by an externally applied
magnetic field.

In Sec. \ref{secTheory}, we presented the details of the theory.  The main results derived in Sec. \ref{secTheory}
include an expression for free energy density $F$ as a function of temperature $T$ and vector potential {\bf a}, and
two basic equations of the theory: the first is an implicit solution for energy gap parameter amplitude $|\Delta_{\bf
k}|$ as a function of wave vector {\bf k}, temperature $T$ and vector potential {\bf a}; and the second is a
London-like relation between electrical current density {\bf j} and vector potential {\bf a}, with an ``effective
superconducting electron density'' $n_s$ that is both $T$- and {\bf a}-dependent. The two equations allow
determination of the spatial variations of {\bf a} and $|\Delta_{\bf k}|$ in a superconductor for given temperature
$T$, applied magnetic field ${\bf H}_a$ and sample geometry. In the low-field limit, the theory reduces to the
local-limit result of BCS.  We also numerically analyzed the functions $|\Delta_{\bf k}(T,{\bf a})|$, $n_s(T,a)$ and
$F_c(T,a)$ (where $F_c$ is the condensation energy density).

In Sec. \ref{secAppl}, as examples, we applied the theory to the case of a semi-infinite superconductor in an applied
magnetic field ${\bf H}_a$ parallel to the surface of the superconductor and the case of an isolated magnetic vortex
in an infinite superconductor, and determined, for each case, spatial variations of quantities such as {\bf a},
$|\Delta_{\bf k}|$, $n_s$ and $F_c$.  We also calculated magnetic field penetration depth $\lambda(T,H_a)$ and lower
critical magnetic field $H_{c1}(T)$.

An important conclusion of the theory is that, depending on temperature $T$, applied magnetic field ${\bf H}_a$ and
sample geometry, a ``partly-paired state'' can exist in which paired electrons (having $|\Delta_{\bf k}|>0$) and
de-paired electrons (having $|\Delta_{\bf k}|=0$) co-exist.  Such a ``partly-paired state'' exists even at $T=0$ when
$H_a$ is above a threshold for a given sample, giving rise to a non-vanishing Knight shift in the superconducting
state at $T=0$ for $H_a$ above the threshold.

Since the non-local effect (or coherence effect) in the superconducting state is not accounted for in the present
theory, we expect the theory to be valid only for highly-local superconductors (for which magnetic field penetration
depth $\lambda$ is much larger than coherence length $\xi$).  However, when a more complete theory is developed that
is able to account for the non-local effect (or coherence effect), we expect it to reduce to the present theory in
the local limit.

\appendix

\section{ Derivation of Eq. (\ref{haoEq}) }
\label{deriveHaoEq}

For ${\bf a}=0$, Eq. (\ref{haoEq}) was previously derived by the author in Ref. \onlinecite{hao93} (related
discussions are also given in Refs. \onlinecite{hao96} and \onlinecite{hao}). The derivation of Eq. (\ref{haoEq}) for
${\bf a}\neq 0$ is similar to that for ${\bf a}=0$. We now present the details of the derivation of Eq.
(\ref{haoEq}).

For convenience, we define
\begin{equation}
\label{Ck} C_{\bf k} = \frac{ 1 - f_{\bf k} - f_{-\bf k} } { 2E_{\bf k}^{(s)} },
\end{equation}
which is a real number.  Then, the self-consistency equation, Eq. (\ref{gap-eq}), can be rewritten as
\begin{equation}
\Delta_{\bf k} = -\sum_{\bf k'} V_{\bf kk'}C_{\bf k'}\Delta_{\bf k'} . \label{aGapEq}
\end{equation}

In the presence of an applied magnetic field, we expect $\Delta_{\bf k}$ to be a function of temperature $T$ and
vector potential ${\bf a}=(a_1, a_2, a_3)$, where $a_i$ ($i=1,2,3$) are the components of {\bf a}. Let $X$ denote any
one of $T$, $a_1$, $a_2$ and $a_3$. We operate $\partial/\partial X$ on both sides of Eq. (\ref{aGapEq}) to obtain
\begin{equation}
\frac{\partial\Delta_{\bf k}}{\partial X} =  -\sum_{\bf k'} V_{{\bf k},{\bf k'}} \left( \frac{\partial C_{\bf
k'}}{\partial X} \Delta_{\bf k'} + C_{\bf k'} \frac{\partial\Delta_{\bf k'}}{\partial X}  \right). \label{partialX}
\end{equation}

We next multiply both sides of the above equation by $C_{\bf k}\Delta_{\bf k}^{\star}$, and then take summation over
{\bf k}, i.e.,
\begin{eqnarray}
\sum_{\bf k} C_{\bf k}\Delta_{\bf k}^{\star}\frac{\partial\Delta_{\bf k}}{\partial X} = \sum_{\bf k'}
\left( - \sum_{\bf k}V_{{\bf k},{\bf k'}} C_{\bf k}\Delta_{\bf k}^{\star} \right) \nonumber \\
\times \left(\frac{\partial C_{\bf k'}}{\partial X} \Delta_{\bf k'} + C_{\bf k'} \frac{\partial \Delta_{\bf
k'}}{\partial X} \right). \label{a3}
\end{eqnarray}

The quantity inside the first pair of parentheses on the right-hand side of the above equation equals to $\Delta_{\bf
k'}^{\star}$ [according to Eq. (\ref{aGapEq})]  so that the second of the two terms on the right-hand side is the
same as the term on the left-hand side.  Thus, we have
\begin{equation}
\sum_{\bf k} |\Delta_{\bf k}|^2 \frac{\partial C_{\bf k}}{\partial X} = 0 \; . \label{aSum}
\end{equation}

We want a $|\Delta_{\bf k}| > 0$ solution. Clearly,
\begin{equation}
C_{\bf k} = \mbox{independent of $T$ and {\bf a}} ,  \label{aHaoEq}
\end{equation}
which is Eq. (\ref{haoEq}) and satisfies
\begin{equation}
\frac{\partial C_{\bf k}}{\partial X} = 0,
\label{partialCk}
\end{equation}
is a solution of Eq. (\ref{aSum}).

However, since Eq. (\ref{aHaoEq}) is not the only possible solution of Eq. (\ref{aSum}) [as one can see, Eq.
(\ref{aSum}) actually can have an infinite number of solutions], we need to justify that Eq. (\ref{aHaoEq}) is the
only physical solution.

Since the diagonalized Hamiltonian, Eq. (\ref{H}), describes a set of independent quasi-particle excitations, there
should be no coupling (except pair correlation) between the quasi-particle excitations.  Therefore, we expect the
thermal energy and entropy associated with each pair of $({\bf k}\!\uparrow,{\bf -k}\!\downarrow)$ excitations to be
\begin{equation}
\varepsilon_{\bf k} = U_{\bf k} + f_{\bf k}E_{\bf k}  + f_{-\bf k}E_{-\bf k}  \label{aek}
\end{equation}
and
\begin{eqnarray}
S_{\bf k} & = & -k_B\left[f_{\bf k}\ln f_{\bf k} + (1-f_{\bf k})\ln (1-f_{\bf k}) \right. \nonumber \\
& & + \left. f_{-\bf k}\ln f_{-\bf k} + (1-f_{-\bf k})\ln (1-f_{-\bf k})\right], \label{aSk}
\end{eqnarray}
respectively.  Similarly, we expect the contribution to the electrical current density from each pair of $({\bf
k}\!\uparrow,{\bf -k}\!\downarrow)$ excitations to be
\begin{equation}
{\bf j}'_{\bf k} = e \left( f_{-\bf k} - f_{\bf k} \right) {\bf v}_{\bf k} \label{ajk}
\end{equation}
[here ${\bf v}_{\bf k}=\hbar{\bf k}/m$ for free electrons, and $\sum_{\bf k}j'_{\bf k}$ corresponds to the second
term on the right-hand side of Eq. (\ref{HaoEq2})].

However, as compared to the above expressions for $\varepsilon_{\bf k}$, $S_{\bf k}$, and ${\bf j}'_{\bf k}$, those
derived from the diagonalized Hamiltonian contain additional terms involving $\partial U_{\bf k}/\partial X$,
$\partial E_{\bf k}/\partial X$ and $\partial f_{\bf k}/\partial X$ (where $X=T$ in the case of $\varepsilon_{\bf k}$
or $S_{\bf k}$; and $X=a_i$ in the case of ${\bf j}'_{\bf k}$). Letting the sum of the additional terms to be zero,
one gets Eq. (\ref{partialCk}), and therefore Eq. (\ref{aHaoEq}).

\section{ Numerical method for solving the boundary-value problem of Sec. \ref{secSemiInfiniteSC}  }
\label{computeAB}

The boundary-value problem of Sec. \ref{secSemiInfiniteSC}, as specified by the system of Eqs. (\ref{bprime}) and
(\ref{aprime}), with boundary conditions given by Eqs. (\ref{bcAt0}) and (\ref{bcAtInfty}), can be solved by using
the Runge-Kutta method.\cite{conte80}

In order to use the Runge-Kutta method, we first need to know values of $a$ and $b$ at one point on the $x$-axis. We
know that deep inside the superconductor both $a$ and $b$ become small, and $n_s$ becomes an $x$-independent
constant, so that the London solutions for $a$ and $b$ hold, which give the following relation between $a$ and $b$:
\begin{equation}
b = \sqrt{n_s} a .
\end{equation}
Let $x_0$ denote the coordinate of such a point located deep inside the superconductor. Since both Eqs.
(\ref{bprime}) and (\ref{aprime}) do not involve coordinate $x$ explicitly, we can assign an arbitrary value to
$x_0$.  We then assign a sufficiently small value for $a(x_0)$, and obtain $b(x_0)$ from the above relation, where
$n_s(x_0)$ is obtained from Eq. (\ref{ns}) for $a=0$ (since the $a$-dependence of $n_s$ is negligible when $a$ is
small) and given $T$.

Once we know $a(x_0)$ and $b(x_0)$ at $x=x_0$, we use the Runge-Kutta method to compute $a(x_n)$ and $b(x_n)$ for
$x_n=x_0-nh$, where $h$ is a small positive interval and $n=1, 2, \cdots, N$, until $b(x_N)\ge H_a$ at $x_N=x_0-Nh$.

Usually, $b(x_N)$ is greater than $H_a$.  If the difference between $b(x_N)$ and $H_a$ is small, we can simply use
$x_N$ as coordinate $x_s$ of the surface of the superconductor. Or, we can obtain $x_s$ by making a linear
interpolation:
\begin{equation}
x_s = x_{N-1} -h \frac{H_a-b(x_{N-1})}{b(x_N)-b(x_{N-1})}.
\end{equation}

\section{ Numerical method for solving the boundary-value problem of Sec. \ref{secSingleVortex} }
\label{computeVortex}

The boundary-value problem of Sec. \ref{secSingleVortex}, as specified by the system of Eqs. (\ref{bprime2}) and
(\ref{aprime2}), with boundary conditions given by Eqs. (\ref{bcAtInfty}) and (\ref{bcFluxQuantum}), can be solved by
using the Runge-Kutta method.\cite{conte80}

In order to use the Runge-Kutta method, we first need to know values of $a$ and $b$ at one point on the $r$-axis.  We
know that far away from the vortex core, both $a$ and $b$ become small and $n_s$ becomes a constant.  For a constant
$n_s$, Eqs. (\ref{bprime2}) and (\ref{aprime2}) can be solved analytically (see, for example, Ref.
\onlinecite{tinkham})):
\begin{equation}
a = \frac{C}{\sqrt{n_s}}K_1(r\sqrt{n_s}) \label{aAtR0}
\end{equation}
and
\begin{equation}
b = CK_0(r\sqrt{n_s}),  \label{bAtR0}
\end{equation}
where $C$ is a constant to be determined, and $K_n(x)$ are the modified Bessel functions of the second kind.

Let $r_0$ denote the $r$-coordinate of such a point located far away from $r=0$, the center of the vortex.  We choose
a value for $r_0$ that is sufficiently large, and guess an initial value for constant $C$, say $C_0$, and obtain
$a(r_0)$ and $b(r_0)$ from Eqs. (\ref{aAtR0}) and (\ref{bAtR0}), respectively, where $n_s$ is obtained from Eq.
(\ref{ns}) for $a=0$ (since the $a$-dependence of $n_s$ is negligible when $a$ is small) and given $T$. Once we know
$a(r_0)$ and $b(r_0)$ at $r=r_0$, the Runge-Kutta method allows us to compute $a(r)$ and $b(r)$ for any $r$.

We then compute total magnetic flux $\Psi$ associated the vortex by numerically carrying out the integral on the
right-hand side of Eq. (\ref{bcFluxQuantum}). If total magnetic flux $\Psi=\Psi_0$ for $C=C_0$ is, for example,
greater than flux quantum $\Phi_0$, we reassign a smaller value for $C$, say $C_1$, and repeat the computation of
$a(r)$, $b(r)$, and $\Psi$.

For the $i$-th repetition ($i\ge 2$), we can assign a value for $C$ by making a linear interpolation or
extrapolation, i.e.,
\[
C_i = C_{i-2} + \frac{\Phi_0 - \Psi_{i-2}}{\Psi_{i-1} - \Psi_{i-2}}\left( C_{i-1}-C_{i-2} \right).
\]

The computation of $a(r)$, $b(r)$ and $\Psi$ is repeated until the difference between total magnetic flux $\Psi$ and
flux quantum $\Phi_0$ is within a predetermined range.  In practice, it usually involves only a few repetitions.

\end{document}